\documentclass[amsmath,twocolumn,superscriptaddress,prb,aps]{revtex4-1} %nofootinbib
\usepackage{amsthm,amsfonts,graphicx,verbatim, color}
\usepackage{bm}% bold math
\usepackage[utf8]{inputenc}
\let\oldmarginpar\marginpar
\renewcommand\marginpar[1]{\-\oldmarginpar[\raggedleft\footnotesize #1]%
{\raggedright\footnotesize #1}}

\newcommand{\be}{\begin{equation}}
\newcommand{\ee}{\end{equation}}
\newcommand{\bea}{\begin{eqnarray}}
\newcommand{\eea}{\end{eqnarray}}

\newcommand{\Tr}{{\rm Tr}\,}
\renewcommand{\epsilon}{\varepsilon}

\renewcommand{\Im}{{\rm Im}\,}

\newcommand{\ket}[1]{|#1\rangle}
\newcommand{\bra}[1]{\langle #1|}
\newcommand{\braket}[2]{\langle #1|#2\rangle}
\newcommand{\tonde}[1]{\left( #1 \right)}

\def\beq{\begin{equation}}
\def\eeq{\end{equation}}
\def\bea{\begin{eqnarray}}
\def\eea{\end{eqnarray}}

\begin{document}

\title{Localized systems coupled to small baths: from A$_{nderson}$ to Z$_{eno}$}
\author{David A. Huse}
\affiliation{Department of Physics, Princeton University, Princeton, New Jersey 08544, USA}
\affiliation{Princeton Center for Theoretical Science, Princeton University, Princeton, New Jersey 08544, USA}
\author{Rahul Nandkishore}
\affiliation{Princeton Center for Theoretical Science, Princeton University, Princeton, New Jersey 08544, USA}
\author{Francesca Pietracaprina}
\author{Valentina Ros}
\affiliation{SISSA, via Bonomea 265, 34136 Trieste, Italy}
\affiliation{INFN, Sezione di Trieste, Via Valerio 2, 34127, Trieste, Italy}
\author{Antonello Scardicchio}
\affiliation{Department of Physics, Princeton University, Princeton, New Jersey 08544, USA}
\affiliation{Princeton Center for Theoretical Science, Princeton University, Princeton, New Jersey 08544, USA}
\affiliation{INFN, Sezione di Trieste, Via Valerio 2, 34127, Trieste, Italy}
\affiliation{Physics Department, Columbia University, New York, NY 10027, USA}
\affiliation{ITS, Graduate Center, City University of New York, New York, NY 10016, USA}
\affiliation{on leave from: Abdus Salam ICTP, Strada Costiera 11, 34151 Trieste, Italy}
\begin{abstract}
We investigate what happens if an Anderson localized system is coupled to a small bath, with a discrete spectrum, when the coupling between system and bath is
specially chosen so as to never localize the bath.  We find that the effect of the bath on localization in the system is a non-monotonic function of the coupling between
system and bath.  At weak couplings, the bath facilitates transport by allowing the system to `borrow' energy from the bath.  But above a certain coupling the bath produces localization, because of an orthogonality catastrophe, whereby the bath `dresses' the system and hence suppresses the hopping matrix element.
We call this last regime the regime of
%, as the coupling to the bath is increased one goes from a localized phase to a weak-localized/delocalized region (depending on the dimensionality of the system) and then back to a new quantum localized phase which we call
``Zeno-localization", since the physics of this regime is akin to the quantum Zeno effect, where frequent measurements of the position of a particle impede its motion. We confirm our results by numerical exact diagonalization.
\end{abstract}
\maketitle

\section{Introduction}
Closed quantum systems can exhibit new dynamical states of matter where they fail to reach thermal equilibrium \cite{Anderson}. Recent years have seen a surge of interest in such many-body localized states of matter
%, because of the advent of many body localization
\cite{Fleishman, agkl, Mirlin, BAA, Oganesyan, Znid, pal}. Quantum localized states exhibit a rich complex of properties, including a vanishing DC conductivity in linear response, a memory of the initial conditions that survives to infinite times in local observables (breakdown of the ergodic hypothesis), an emergent integrability \cite{lbits, serbyn, imbrie, rms-IOM, chandran,kim}, a non-local response to local perturbations \cite{adiabaticity}, and a stabilization of exotic correlated states of matter at high temperatures \cite{LPQO, Pekker, Vosk, Kjall, deluca2013,Bauer, Bahri, lspt, qhmbl} (for a review of recent developments, see Ref. \onlinecite{arcmp}). Quantum localization has been drawing intense interest both because it represents an unexplored frontier for quantum statistical mechanics, and because it holds out the promise of a new generation of quantum devices, that are protected against decoherence and can operate even at high energy densities. However, much work on quantum localization considers only the idealized (and experimentally unrealizable) limit of a completely closed quantum system, perfectly isolated from any environment.

A recent series of works (involving some of the present authors) have studied what happens when a localized system is coupled to a thermodynamically large bath \cite{ngh, johri, gn, 2dcontinuum}. These works have shown that when a localized system is weakly coupled to a large bath, the exact eigenstates of the combined system and bath immediately become thermal, while the spectral functions of local operators continue to show signatures of localization up to a crossover coupling that is independent of the size of the bath.  In the present paper, we instead address what happens if a localized system is exposed to a {\it small} bath, containing very few degrees of freedom. Additionally, we do not want to restrict ourselves to the regime of weak coupling. If a handful of `delocalized' degrees of freedom are exposed to a strongly disordered (localized) system, the most likely result is that these additional degrees of freedom will also become localized.  That is not the physics we consider here.  We want to ask: how many degrees of freedom do we need to have in a bath \emph{that is protected against localization}, in order to be able to thermalize a localized system? We note that baths that are `protected' against localization are not unphysical. Examples include the longest wavelength Goldstone modes (e.g.\ phonons) associated with the spontaneous breaking of a continuous symmetry\cite{Chalker}, as well as extended states in systems with a topological obstruction to the construction of fully localized Wannier orbitals \cite{Thouless}.

In this article, we will examine what happens when a localized system is coupled (potentially strongly) to a {\it small} bath, which is protected against localization. For simplicity, we will restrict our attention to single-particle localized systems. We will show that one and two dimensional single particle localized systems coupled to a {\it finite} sized bath are always localized, irrespective of the strength of the coupling. However, the localization length and inverse participation ratios display a non-monotonic dependence on the coupling between system and bath, which is associated with a crossover between Anderson localization and a regime that we dub `quantum Zeno localized,' where repeated `measurements' of the particle by the bath are responsible for localization. In three dimensions, a delocalized phase can arise at intermediate couplings.

%Although we will give a rough description of the Hamiltonian of the bath in terms of a random matrix, we will ensure that the bath is not localized by choosing the coupling between system and bath so that it does not change the statistics of the bath spectrum.

\section{The model}

We begin by considering as our localized system a single particle %(a fermion for specificity) 
moving on a lattice with a random potential. The Hamiltonian of the particle is:
\begin{equation}
H_0 = -t\sum_{\langle ij \rangle} c^{\dag}_i c_j + \sum_i\epsilon_i c^{\dag}_i c_i
\end{equation}
where $\epsilon_i$ is a random onsite energy taken from a distribution of width $2W$ (specifically, a box distribution $[-W,W]$). The lattice dimensionality is $d$, arbitrary, and the most important difference is between the cases $d=1,2$, where delocalization is impossible and $d\geq 3$ where we could actually have delocalized states.  When we consider a finite lattice, it has $L^d$ sites.

The bath in question is modeled as a quantum dot, or a zero-dimensional system. A quantum dot is a suitable model for the bath, because we want to couple the system to the bath in such a way that the coupling does not introduce any spatial disorder in the bath i.e.\ the system couples uniformly to the entire bath, which means that from the system's point of view, the bath is zero dimensional. We further assume that the bath has bandwidth $\Omega$, and can be in any one of $N$ possible states, so that the level spacing in the bath is $\delta \approx \Omega/N $. In the limit $N \rightarrow \infty$ the bath can have a continuum spectrum. The Hamiltonian of the bath may then be modeled simply as a properly rescaled $N\times N$ Hermitian random matrix taken from the GOE ensemble. The GOE statistics of the bath
%, and in particular level repulsion alone 
is representative of it being in a \emph{delocalized} phase.

The Hamiltonian for the bath is taken to have the form
\begin{equation}
H_{bath}=\omega \sum_{\alpha',\beta'}M_{\alpha',\beta'}\ket{\alpha'}\bra{\beta'},
\end{equation}
where for simplicity of notation we define the variable
$\omega={\Omega}/{2\sqrt{2N}},$ and where $M$ is a GOE matrix distributed according to:
$P(M)\propto e^{-\frac{1}{2}\Tr M^2}$
with
\begin{equation}
\langle M_{\alpha,\beta} \rangle=0, \quad
\langle M_{\alpha,\beta}^2 \rangle=1/2\text{ for }\alpha\neq\beta, \quad
\langle M_{\alpha,\alpha}^2 \rangle=1.
\end{equation}

%If we want to have a bandwidth $\Omega$ for the bath (this is an independent parameter), using the known results for the GOE matrices, we need to include the following term in the Hamiltonian:

The eigenvectors of the bath are labeled by $\{\alpha\}_{\alpha=1,...,N}$. The density of levels $E_\alpha$ is given by the semicircle law \cite{Mehta}
\begin{equation}
\rho(E)=\frac{8 N}{\pi\Omega^2}\sqrt{(\Omega/2)^2-E^2},
\end{equation}
and hence $\delta\equiv 1/\rho(0)=\pi\omega/ \sqrt{2 N}$ in the middle of the spectrum.

The coupling between system and bath is chosen so as to not introduce localization into the bath, but also so that it is able to transfer energy between system and bath. The simplest coupling that does the job is
\begin{equation}
H_{couple} = \lambda \sum_{i, \alpha, \beta} M^{(i)}_{\alpha,\beta}\ c^{\dag}_i c_i\otimes |\alpha\rangle \langle \beta| \label{couple}
\end{equation}
i.e. a coupling of strength $\lambda$ which can scatter the bath from any eigenstate to any other eigenstate (irrespective of the energy transfer involved) with a random amplitude $M^{(i)}_{\alpha,\beta}$. For simplicity, we choose the amplitudes to form a random GOE matrix.

% [I covered this para, since I don't understand it and it does not seem important.  ---DH]  We note that the question of whether the bath is interacting or not has not entered. Such details can strongly influence how $\delta$ scales with the size of the bath, but for our purposes the bath may be treated as a black box, and the only quantity that will matter for the analysis is the accessible level spacing $\delta \sim \omega / \sqrt{N}$.

We now ask what happens to the particle in the presence of this coupling to the bath. It is essential for our present purposes that the system contains a
single particle (if the system contained many particles we would have to worry about indirect couplings through the bath.) The hopping problem can be
pictorially represented as in Fig. 1. For every position of the particle, there is a `tower' of $N$ states, which differs only in the configuration of the bath.
This `tower' of states has bandwidth $\sim \omega \sqrt{N}$ and level spacing $\sim \omega / \sqrt{N}$. A nearest neighbor hop of the particle, leaving the state of
the bath unchanged, causes an energy shift of magnitude $W$ (in the weak $\lambda$ limit). For the rest of the paper we assume that
$\omega/\sqrt{N}<W < \omega \sqrt{N}$, so that the `offset' of the tower of states on neighboring sites is bigger than the level spacing in the tower,
but nonetheless adjacent towers do overlap.  Right at the edge of the towers of states there are Lifshitz tails - states that are not near degenerate with
any nearby states.  However, we consider the properties of typical states well away from the edges of the spectrum, where the towers of states all overlap.
%Not surprisingly, this is the most interesting region of parameters.

The single particle problem will be considered in three stages. First we consider what happens working perturbatively in small $\lambda < \omega / \sqrt{N}$. Next we discuss the regime of strong $\lambda > \omega$. Finally, we consider the intermediate $\lambda$ regime $\omega / \sqrt{N} < \lambda < \omega$.

\section{Anderson and Zeno localization}

\subsection{Weak $\lambda < \omega / \sqrt{N}$: the Anderson localized regime}
\begin{figure}
\includegraphics[width = \columnwidth]{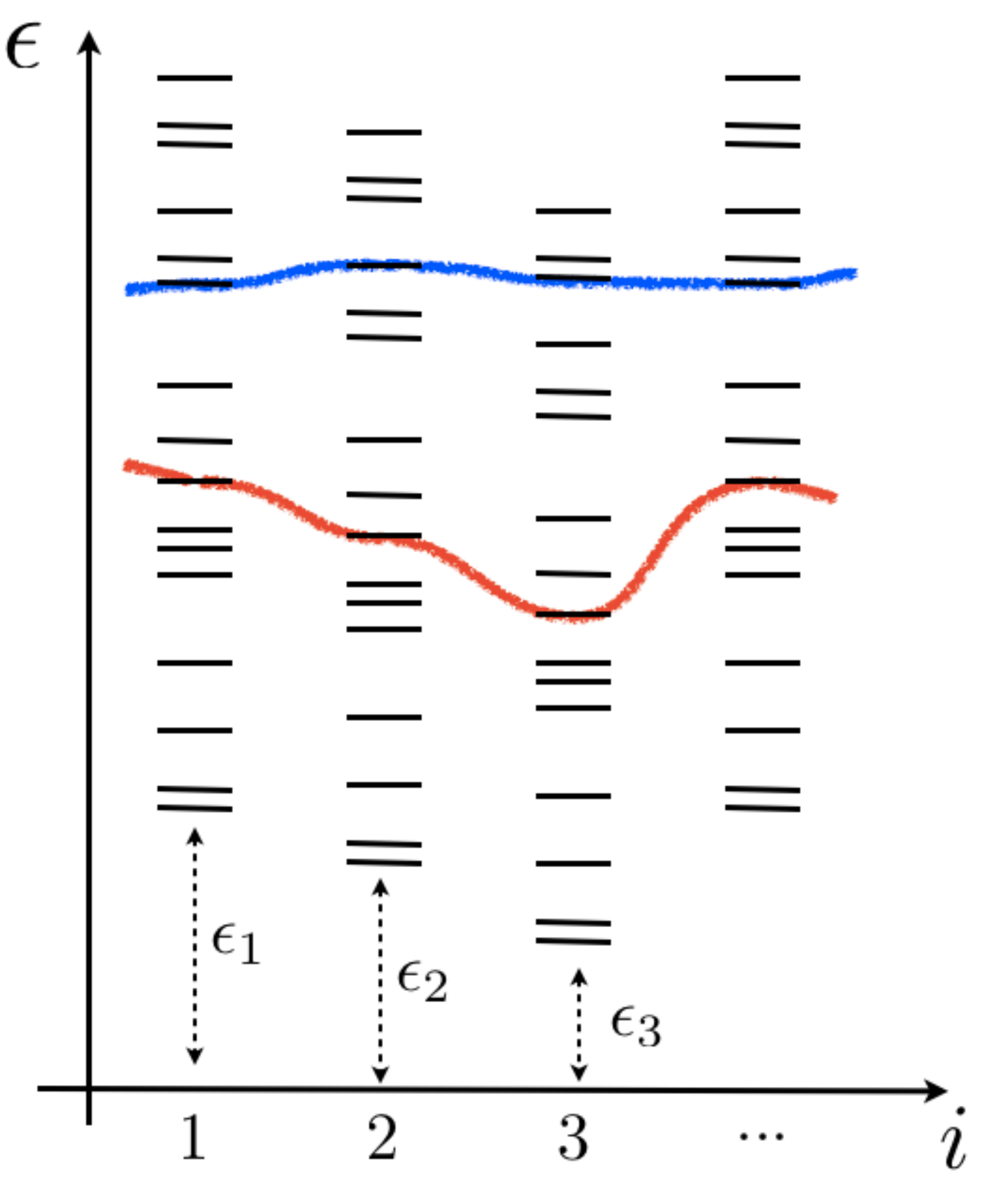}
\caption{Figure illustrating the basic setup. There is a band of states on every site $i$, with energies spanning a bandwidth $\omega \sqrt{N}$, and with level spacing $\omega/\sqrt{N}$, which differ only in the state of the bath. In the weak coupling regime $\lambda < \omega/\sqrt{N}$, hopping between sites follows the solid red lines (with no change in the state of the bath), and a typical nearest neighbor hop changes the energy by an amount of order $W$. Outside the weak coupling regime, the states in the bath start to get hybridized. The size of the energy window over which states in the bath are well hybridized grows with $\lambda$, and becomes of order $W$ at $\lambda \sim \sqrt{\omega W}/ N^{1/4}$.
%At this point, there arises an 
The system-bath coupling allows an effective hopping between states that are nearly on shell (blue lines).
The system is minimally localized (for small $t$) at $\lambda \sim \sqrt{\omega W}/ N^{1/4}$, with larger values of $\lambda$ leading to greater localization because of the quantum Zeno effect.  }
\end{figure}
At $t=0=\lambda$, the spectrum of the combined system and bath consists of decoupled `bands' of states (one band per site), with bandwidth $\omega \sqrt{N}$ (Fig. 1) and random offsets $\epsilon_i$ (see Fig. 1). Our assumption that $W\ll \Omega=2\omega \sqrt{2 N}$ ensures that these bands overlap strongly.
 %As we are working at infinite temperature, this is the region of the spectrum we are interested in.

Although the statistics of the bath alone are Wigner-Dyson, the overall spectral statistics are Poisson. This is because there are $L^d$ local integrals of motion $c^\dag_i c_i$, which commute with the Hamiltonian, so the spectrum is the superposition of $L^d$ copies of $H_{bath}$ spectra, shifted by the random energies $\epsilon_i$. Meanwhile, the eigenstates take the form\begin{equation}
\ket{\Psi}=\ket{i}\otimes\ket{\alpha},
\end{equation}
where $\ket{\alpha}$ is an eigenstate of the bath Hamiltonian with energy $E_{\alpha}$.

 On turning on non-zero $t$ (but still at $\lambda = 0$), the system becomes able to execute hopping from one site to the next, but the hopping does not involve any change in the state of the bath, and thus a nearest neighbor hop typically involves an energy change of order $W$. The eigenstates are still `product states' of system and bath. For $t < W$ and $\lambda =0$ we are in the regime of strong Anderson localization, where the localization length is less than or of order one lattice spacing, and the `system' part of the eigenstate is just a dressed version of $\ket{i}$. In this regime we can apply a variant of the locator expansion, a perturbation theory in the hopping or interaction \cite{agkl} recently also used in the context of MBL \cite{rms-IOM, laumann2014many}. For $t > W$ the traditional locator expansion will fail to converge, and we will be in either a weak localization regime (in one or two dimensions), or a delocalized regime (in three dimensions). As long as the particle is localized, there remain $L^d$ local integrals of motion (the occupation numbers of the localized eigenfunctions), and the spectral statistics thus remain Poisson  \footnote{In the following we will consider states in the middle of the spectrum. Outside of the center of the band other phenomena arise. In particular due to the reduction of the density of states the localized phase is expected to be larger than the limits computed here.}.

We now move to non-zero $\lambda$. %Upon turning on $\lambda,$ the first thing that happens is that the bath states start mixing. For the time being we will treat $t$ perturbatively (assuming $t\ll W$) -- we will comment on the weakly localized/delocalized region later on.
When the particle is at position $i$, the effective Hamiltonian for the bath is $ \omega M^{(0)}+\lambda M^{(i)}$. The `bath' eigenstates will start to mix with each other when $\lambda$ becomes comparable to the level spacing $\sim \omega /\sqrt{N}$ associated with the bare bath Hamiltonian $M^{(0)}$. A more careful argument along the lines of Ref. \onlinecite{agkl,de2014anderson} interprets the above expression as the Hamiltonian of a fully connected graph, with connectivity $N$, random on-site energies (eigenvalues of $M^{(0)}$) and random Gaussian hopping $\lambda M^{(i)}_{\alpha,\beta}$. This leads to a slightly more precise criterion, stating that the eigenstates of the bath must remain almost unperturbed for $ \lambda\lesssim \omega/\sqrt{N} \ln N$. However, this fine distinction is of little importance for the present analysis (we are considering small baths).

Therefore for $\lambda<\omega / \sqrt{N}$, hopping between neighboring sites is not enhanced by the presence of the bath.
In fact, let us consider the $O(t)$ correction to the localized eigenstate
\begin{equation}
\ket{\Psi}\simeq \ket{i}\otimes\ket{\alpha}+\sum_{\beta}A_{i+1,\beta}\ket{i+1}\otimes\ket{\beta}+...\ .
\end{equation}
If $\lambda<\delta$, for hopping leaving the bath untouched, perturbation theory gives
\begin{equation}\label{hop1}
A_{i+1,\alpha}=\frac{t}{\epsilon_{i+1}-\epsilon_{i}}\sim \frac{t}{W},
\end{equation}
while for hopping that changes the state of the bath $\beta\neq\alpha$ the same ratio is at most
\begin{eqnarray}\label{hop2}
\max_\beta A_{i+1,\beta}&=&
\frac{t}{\epsilon_{i+1}-\epsilon_{i}}\max_{\beta}\frac{\lambda}{\epsilon_i-\epsilon_{i+1}+E_\alpha-E_\beta}\nonumber\\
&\sim&\frac{t}{\epsilon_i-\epsilon_{i+1}}\frac{\lambda}{\delta} \sim \frac{t \lambda}{W \delta}<A_{i+1,\alpha}.
\end{eqnarray}
So in this regime, the bath is typically not excited by the particle traveling. This is illustrated in Fig. 1: the solid red lines indicate the trajectory followed by a particle hopping without changing the state of the bath.

We now consider how the criterion for breakdown of the locator expansion ($t > W$ for $\lambda = 0$) is altered at non-zero $\lambda$. We recall that in the
regime $\lambda < \omega/\sqrt{N}$, the bath does not respond to the motion of the particle in the system. However, the coupling of the bath to the system
changes depending on where the particle is (Eq. \eqref{couple}). Thus, in this weak coupling regime the bath acts as an additional source of static disorder.
The particle is effectively hopping in a random potential with disorder strength $W_{eff} \sim \sqrt{W^2 + \lambda^2} \approx W (1+\lambda^2/2W^2)$.
Meanwhile, from Eqs. \eqref{hop1},\eqref{hop2} we see that the bath opens up additional hopping channels, and increases the effective hopping from $t$ to $t_{eff} \approx t (1+ \lambda^2/2 \delta^2)$. The condition for the breakdown of the locator expansion is altered to $t_{eff} > W_{eff}$. Since $\delta < W$, we see that the opening up of new hopping channels is the dominant effect, and thus coupling to a bath makes localization less stable, changing the critical hopping to $t_c=W(1-c\lambda^2/W^2)$ (in high dimensions the result is modified as an extra factor is needed, to get $t_c=W(1-c\lambda^2/W^2)/(d\ln d)$). For $t < t_c$, we have strong localization, and for $t > t_c$ we have either weak localization (in one or two dimensions) or delocalization (in three dimensions). We note too that in the weakly coupled strong localization regime the exact eigenstates are effectively product states of system and bath, and the entropy of entanglement of the system with the bath is near zero.
%\begin{equation}
%\ket{\Psi}=\sum_{j}A_{j,\alpha}\ket{j}\otimes\ket{\alpha}=\ket{\psi}\ket{\alpha},
%\end{equation}
%where $A_{j,\alpha}\sim (t/W)^{|j-i|}$, where $i$ is the center of localization.

%\emph{Entanglement.}
%In this case the bath remains in the state $\alpha$ and the eigenstates are of the form
%\begin{equation}
%\ket{E}=\ket{\phi}\ket{\alpha}
%\end{equation}
%where $\phi$ is one of the localized states and $\alpha$ is a state from the bath. The entanglement between bath and particle in this case is null:
%\begin{equation}
%\rho_{i,i'}=\Tr_{Bath}(\ket{\phi}\ket{\alpha}\bra{\alpha}\bra{\phi})=\phi^*_i\phi_{i'}
%\end{equation}
%and
%\begin{equation}
%S=-\Tr\rho\ln\rho=0.
%\end{equation}
%

\subsection{ Strong $\lambda > \omega $: The quantum Zeno regime} In the opposite limit of strong $\lambda$, the particle gets localized again, because of the coupling to the bath. We dub this the regime of quantum Zeno localization, because of the resemblance to the Quantum Zeno effect \cite{Beskow, Khalfin, Misra} : the fact that a small system coupled with a large quantum system, possibly a detection apparatus, does not evolve or evolves only into a given subspace \cite{Facchi2001,Facchi}, when the coupling is too large.

The calculation proceeds as follows: first, we observe that at $t=0$, a $\lambda>\delta$ causes a hybridization of the levels in the bath which now acquire an $i$ index: 
\begin{equation}
\left(\omega M^{(0)}+\lambda M^{(i)}\right)\ket{\alpha_i}=E_{\alpha_i}\ket{\alpha_i}.
\end{equation}

For $\lambda \gg \omega$, the Hamiltonian of the bath is dominated by the coupling $M^{(i)}$ to the particle, and the bath levels are hybridized in a radically different way for each position $i$ of the particle. Thus we have
\begin{equation}
\braket{\alpha_i}{\beta_j}=\delta_{\alpha,\beta} \delta_{i,j} + (1-\delta_{i,j}) x_{ij}/\sqrt{N}, \label{eq: hybridization}
\end{equation}
where the $\delta_{i,j}$ is a Kronecker delta function and $x_{ij}$ is a Gaussian random variable $<x_{ij}>=0$ and $<x_{ij}^2>=1$.

We now turn on a small $t$ and ask how the analysis changes. A hop in the system changes the state in the bath. We can describe the problem by mapping it to a Bethe lattice problem with effective hopping
\begin{equation}\label{p1}
\tau=t\frac{1}{\sqrt{N}},
\end{equation}
effective disorder
\begin{equation}
\mathcal{W}=\lambda\sqrt{N}
\end{equation}
and connectivity $\kappa=N$. Using the known results on the localization on Bethe lattice we have localized eigenstates if
\begin{equation}\label{eq:BL}
t\lesssim\lambda/\ln N.
\end{equation}
For such values of $t$ the particle is strongly localized because of quantum Zeno physics, whereas for $ t > \lambda/\ln N$ the locator expansion fails to converge. This latter regime may be either a weak localized regime (in one or two dimensions) or a delocalized regime (in three dimensions). The localization of the particle by strong $\lambda$ can also be viewed as a result of an orthogonality catastrophe, whereby the particle is `dressed' by the bath in a different way depending on which site it is on, and the hopping matrix element is thus strongly suppressed. In the limit $\lambda \rightarrow \infty$ the hopping is completely ineffective, and the exact eigenstates are simply product states $\ket{\Psi}=\ket{i}\ket{\alpha_i}$, which, however, are exact eigenstates of the system-bath coupling. In this limit, the entropy of entanglement of system and bath (in an eigenstate) is again zero, and the particle is localized on a single site.

\subsection{Intermediate $ \omega / \sqrt{N}<  \lambda < \omega $}
We begin our discussion of the intermediate coupling regime by setting $t = 0$, and studying the evolution of the bath eigenstates as $\lambda$ is varied. Turning on a coupling $\lambda > \omega / \sqrt{N}$ causes the eigenstates of $M^{(0)}$ within an energy window $\Delta$ to hybridize.  The width of this energy window may be determined by calculating the decay rate of an eigenstate of $\omega M^{(0)}$ due to the perturbation $V=\lambda M^{(i)}$ using Fermi's golden rule. Given a density of final states $\delta^{-1}=\sqrt{2 N}/\pi \omega$ the calculation indicates that the decay rate is
\begin{equation}
 \Delta\simeq 2\pi\lambda^2 \sqrt{2N}/ \pi \omega, \label{Delta}
 \end{equation}
and moreover suggests that the broadened spectral line is a Lorentzian with width $ \Delta$. Thus, the eigenstates $\ket{\alpha_{(i)} }$ of $\omega M^{(0)} + \lambda M^{(i)}$ should be wave packets of eigenstates $\ket{\alpha_{(0)}}$ of $\omega M^{(0)}$, with
\begin{equation}\label{A}
|\bra{\alpha_{i}}\ket{\alpha_{0}}|=\sqrt{\frac{\delta\Delta / \pi}{(E_{\alpha_{i}}-E_{\alpha_{0}})^2 + \Delta^2}}.
\end{equation}
%(the phase is irrelevant for the following reasoning as no interference occurs).
We note that as $\lambda \rightarrow \omega$, $\Delta \rightarrow 2\omega \sqrt{2N}=\Omega$ indicating complete hybridization of all states, whereas as $\lambda \rightarrow \delta$, $\Delta \rightarrow 2 \pi \delta$, indicating no hybridization. Thus the Fermi's golden rule interpolation correctly matches on the weak and strong $\lambda$ limits. The line broadening $\Delta$ becomes comparable to $W$ for
\begin{equation}
\lambda_c = \sqrt{\frac{{W \omega}}{2 \sqrt{2 N}}}.
\label{lambdac}
\end{equation}
We now turn on a small but non-zero $t$. Hopping is perturbative in $t$ but non-perturbative in $\lambda$ and the eigenstates have the form 
\begin{eqnarray}
\ket{\Psi}&=&\ket{i}\ket{\alpha_{i}}\nonumber\\
+&\sum_{\alpha_{i+1}}&\frac{t\bra{\alpha_{i+1}}\ket{\alpha_{i}}}{\epsilon_{i+1}+E_{\alpha_{i+1}}-\epsilon_{i}-E_{\alpha_{i}}}\ket{i+1}\ket{\alpha_{i+1}}\nonumber\\
&+&...\ .
\end{eqnarray}
%Now the minimum of $\epsilon_i+E_{\alpha_{i}}-\epsilon_{i+1}-E_{\alpha_{i+1}}\sim\delta=\omega/\sqrt{N}$, and
A `direct' hopping (blue line in Fig. 1), which stays on shell to a precision $\delta \sim \omega/\sqrt{N}$, is now possible, because there is some non-zero overlap between the two states $\bra{\alpha_{i+1}}$ and $\ket{\alpha_{i}}$ at the same energy. However, since $|\epsilon_i - \epsilon_{i+1}| \sim W$, these `direct hopping' processes must involve transitions between bath states with $|E_{\alpha_{i}} - E_{\alpha_{i+1}}| \sim W$. The amplitude of the overlap between bath states may be estimated by inserting  $(E_{\alpha_{i}}-E_{\alpha_{0}}) \sim W$ into Eq. \eqref{A}. (Identical results may be obtained by instead using the Golden Rule to calculate the decay rate of $\ket{\alpha_{i}}$ onto eigenstates of $\omega M^{(0)} + \lambda M^{(i+1)}$).  Thus the correction to the wave function from `direct hopping' processes (blue line in Fig. 1) is, at leading order in small $t$, %
\begin{equation}
\sim\left(\frac{t}{\delta} \frac{\sqrt{\Delta\delta}}{\sqrt{W^2 + \Delta^2}}\right).
\end{equation}
For $\lambda > \lambda_c$, we have $\Delta > W$ and the above expression can be approximated by $\frac{t }{\lambda}$ (remembering that $\Delta \sim \lambda^2/\delta$). The same result can also be obtained by reasoning that for $\lambda > \lambda_c$, direct hopping via the blue line in Fig. 1 is `easy', since bath states are hybridized over an energy window $\Delta > W$. However, the matrix elements are suppressed by a factor of $\sqrt{\tilde N}$, where $\tilde N = \Delta/\delta \sim (\lambda/\omega)^2 N$  is equal to the number of states involved in the hybridization (for $\lambda>\omega$, $\tilde N=N$).

Meanwhile, for $\lambda < \lambda_c$, we have $\Delta < W$, and the above expression can be approximated by ${t \lambda}/{W \delta}$. This expression may be understood as follows: since $\Delta < W$, a `direct hop' (following the blue line in Fig. 1) is forbidden, as the two bath states involved have vanishing overlap. Instead, the particle first hops without changing the state of the bath, going off shell by an amount $W$, and then the bath relaxes to bring the system back on shell, to a precision $\delta$. The matrix element for this two step process is $t \lambda / W$.

We have thus shown that when performing a locator expansion in small $t$, the successive corrections to the wave function are suppressed by powers of $t/\lambda$ if $\lambda > \lambda_c$, and by powers of $t \lambda / W \delta$ if $\lambda < \lambda_c$ and $\lambda>\delta$.

Thus, summarizing, the locator expansion converges for
\begin{equation}
\label{convergence}
t\lesssim \begin{cases}
W -c\lambda^2/W &\text{ if } \lambda<\delta\\
W\delta/\lambda &\text{ if } \delta<\lambda<\lambda_c\\
\lambda/\ln N &\text{ if } \lambda>\lambda_c.
\end{cases}
\end{equation}

This behavior is non-monotonic. For $\lambda < \lambda_c$, a stronger coupling to the bath destabilizes localization, by allowing the system to `borrow' energy from the bath to hop, whereas for $\lambda > \lambda_c$ a stronger coupling to the bath stabilizes localization. This latter effect is the result of an orthogonality catastrophe - a stronger $\lambda$ suppresses the effective hopping matrix element, because it suppresses the overlap between bath states corresponding to the particle being on different sites.

\subsection{Region of convergence of the locator expansion}
\begin{figure}
\includegraphics[width = \columnwidth]{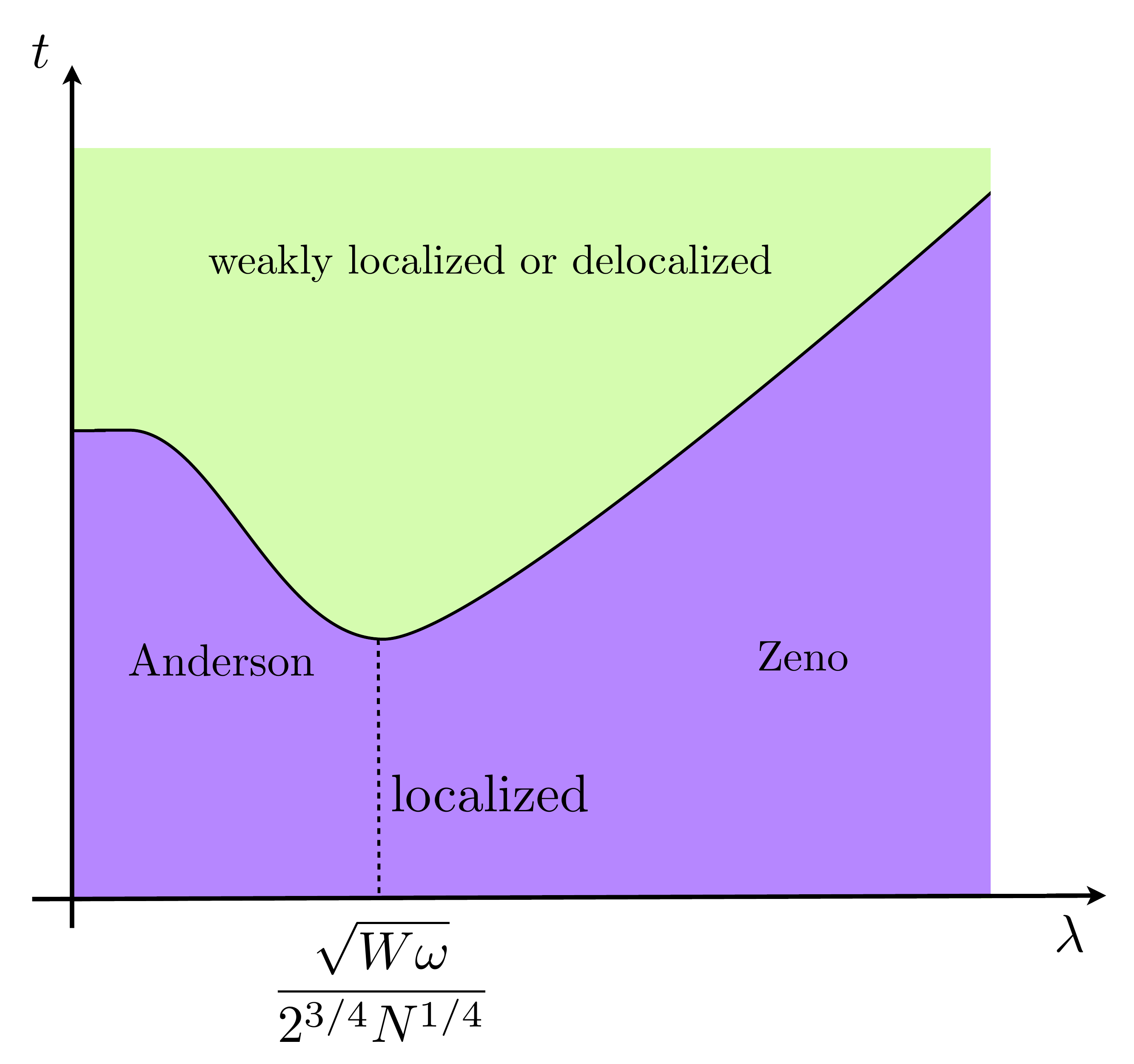}
\caption{\label{fig:phdiag} Schematic phase diagram, with a `phase boundary' that indicates the boundary of stability of the locator expansion. On the small $t$ side of the `phase boundary,' the system is strongly localized. Within the strong localization regime there is a crossover from Anderson localization at small $\lambda$ to quantum Zeno localization at large $\lambda$. Note that the locator expansion is maximally unstable around $\lambda_c = \sqrt{\omega W/(2\sqrt{2N})}$. For $\lambda > \lambda_c$, the phase boundary is approximately linear in $\lambda$, whereas for $\lambda < \lambda_c$ is scales as $1/\lambda$. Meanwhile, the large $t$ side of the `phase boundary' is the regime of weak localization (in one or two dimensions) or delocalization (in three dimensions). Note that in one or two dimensions the `phase boundary' marking breakdown of the locator expansion is really just a crossover to weak localization. Only in three dimensions does it mark a true phase transition to a delocalized phase. }%   strong localization, quantum Zeno localization, and delocalization/weak localization respectively, whose properties are discussed more fully in the main text. The boundaries of the various regimes are crossovers in one and two dimensions, but in three dimensions there is a phase transition separating the delocalized regime from the localized regimes.}
\end{figure}

The results from the previous three subsections can be summarized by Fig. \ref{fig:phdiag}. At the smallest $\lambda < \omega / \sqrt{N}$, the coupling to the bath destabilizes localization by opening up new hopping channels, and the critical hopping scales as $t_c \sim W (1- c\lambda^2/W^2)$, where $c$ is an $O(1)$ numerical prefactor. For $\omega / \sqrt{N} < \lambda < \lambda_c = \sqrt{\omega W/(2\sqrt{2N})}$, the coupling to the bath assists the particle in hopping, by allowing it to `borrow' the energy required to get on shell. For $\lambda > \lambda_c =  \sqrt{\omega W/(2\sqrt{2N})}$, the coupling to the bath again enhances the stability of localization, because of a `quantum Zeno effect.' %The region of convergence of the locator expansion can be estimated as $t < \sqrt{W^2 + \lambda^2}$ for $\lambda <  \delta = \omega / \sqrt{N}$, as Eq. \ref{convergence} for $\omega/\sqrt{N} < \lambda < \omega $, and as $t < \lambda$ for $\lambda > \omega$. 
As long as the locator expansion converges, the system will be in a `strong localization' regime, with $\lambda$ driving a crossover from `Anderson' localization (at weak $\lambda$) to `Zeno' localization (at strong $\lambda$). What happens when the locator expansion fails to converge depends on dimensionality, and will be discussed in the following section.

An essentially similar but more quantitatively precise calculation of the boundary of stability of the locator expansion is provided in the Appendix, and leads to Fig. \ref{fig:analytic}. We outline here the main calculation. For a general eigenstate, we consider the amplitude $A_{j,\beta}$ (defined as in Section IIIC) for the particle to be in the site $j$ of the lattice, with the bath being in the state $\ket{\beta}$.

For a state localized in the vicinity of a site $i$, the amplitude to find a particle at a site $j$  at distance $n$ from $i$ is exponentially small in the distance, implying that for some $z<1$  \begin{equation}
P\left(\max_\beta|A_{j,\beta}|<z^n\right)\to 1,
\end{equation}
as $n\to\infty$, where $P$ is the probability measure over the realizations of the disorder. The minimum $z$ for which this condition is still true gives the localization length as $z=e^{-a/\xi}$ where $a$ is the lattice constant.

Analytic calculations, that can be done by considering $A_{j,\beta}$ to lowest order in $t$ in perturbation theory (see also Ref.s \cite{agkl,de2014anderson,rms-IOM,laumann2014many}) are presented in the Appendix. Being performed in the lowest-order in $t$ these give a lower bound for the value of $t$ where delocalization/weak-localization occurs.

So we conclude that the boundary of convergence for a locator expansion should look as sketched in Fig. \ref{fig:phdiag} and \ref{fig:analytic}. Note that strong localization is least stable when $\lambda \approx \lambda_c$, and becomes more stable both for weak $\lambda$ (the Anderson localization limit), and for strong $\lambda$ (the quantum Zeno limit). The minimum value of $t$ that can cause breakdown of the locator expansion is $t_c(\lambda_c) = \sqrt{\omega W/(2\sqrt{2N})}$. So, for any $t$ the localization length should peak at this value of $\lambda$. This is observed both in the numerics (Fig.4) and in the analytic calculations (Fig. \ref{fig:analytic}).

\begin{figure}[htbp]
\begin{center}
\includegraphics[width=\columnwidth]{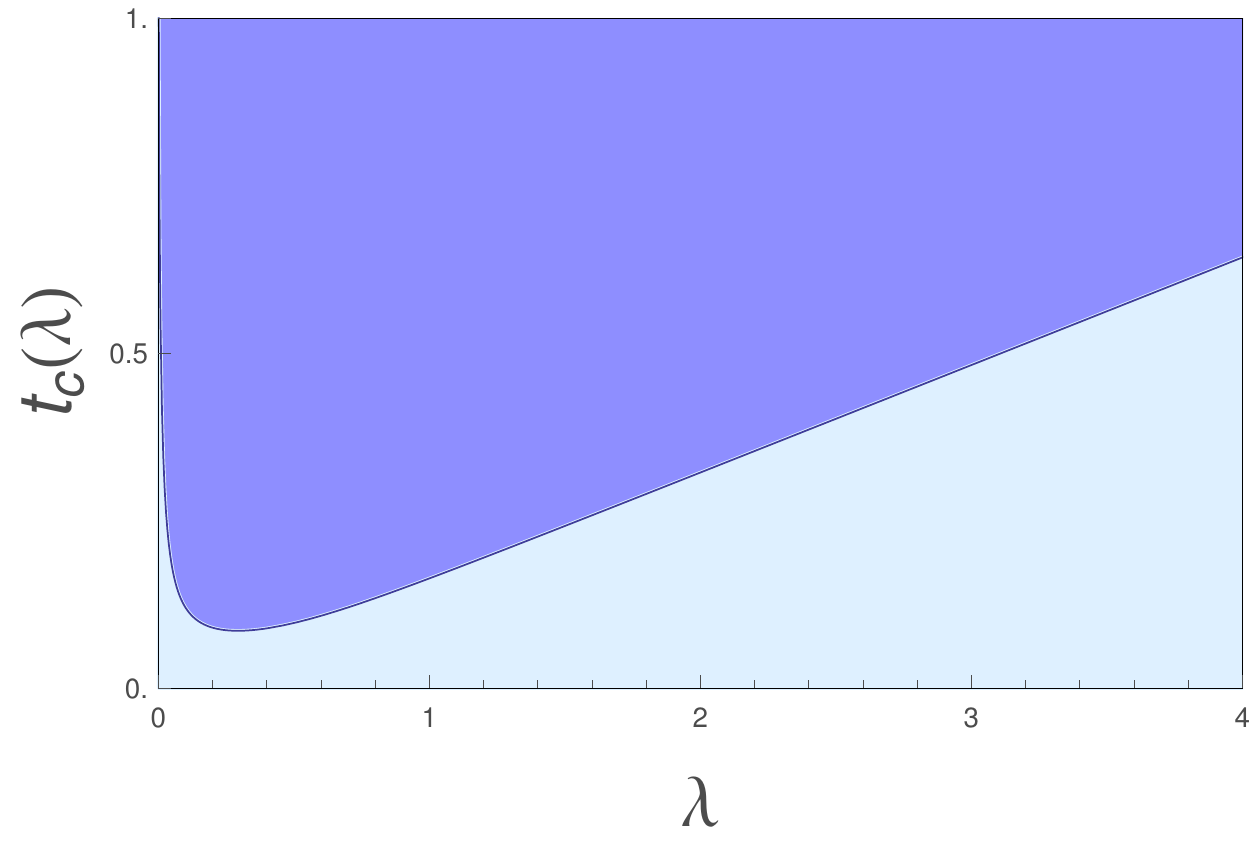}
\includegraphics[width=\columnwidth]{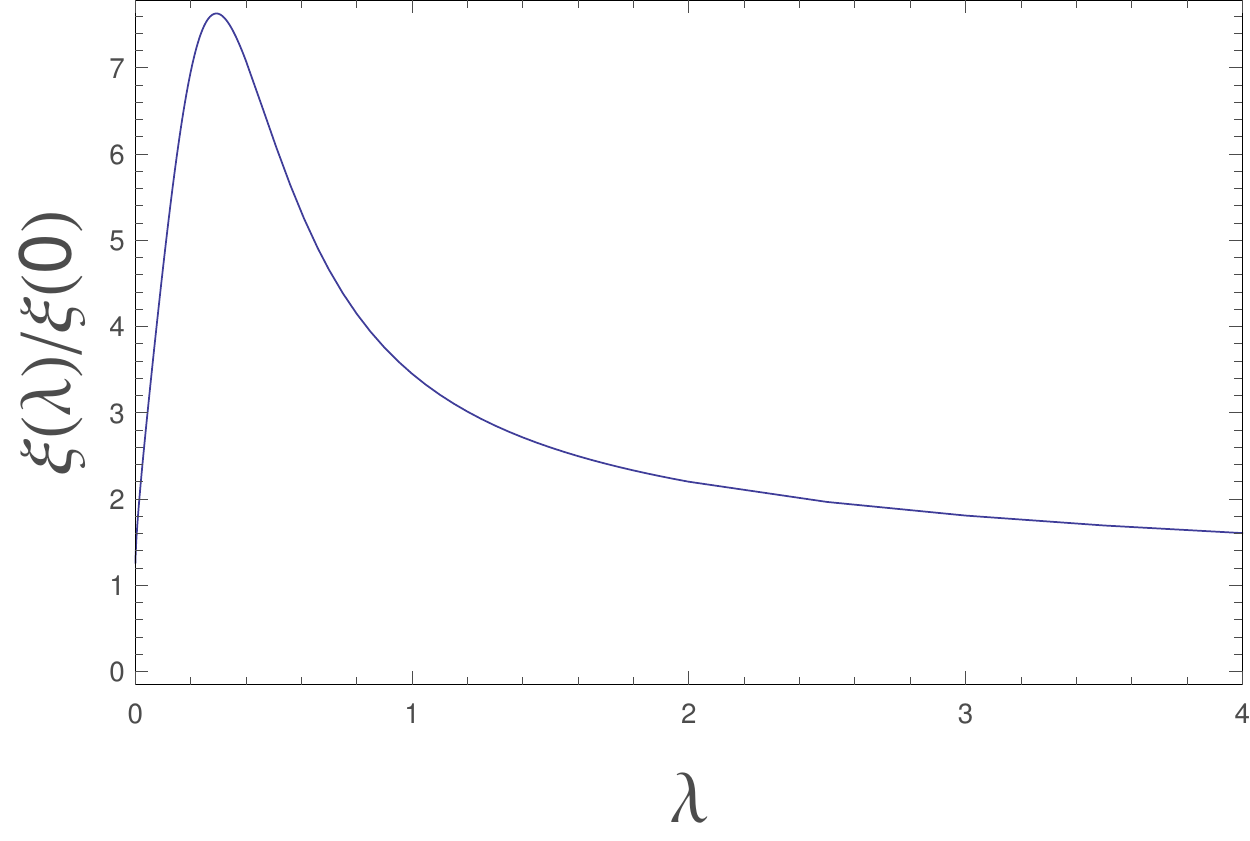}
\caption{Results of analytic calculations in the forward approximation detailed in the Appendix for fixed $W=3$, $N=300$, $\omega=4$ (substitution into Eq. \eqref{lambdac} gives $\lambda_c = 0.5$).  {\it Top.} The $t-\lambda$ phase diagram; the dark region is the weak-localized/delocalized region, the light region is localized. Note the non-monotonic dependence of the boundary on $\lambda$. The calculation assumes that states in the bath are hybridized according to Eq. \eqref{A}, and is thus not applicable at $\lambda < \delta$. {\it Bottom.} Localization length as a function of $\lambda$, along a horizontal slice through the top diagram that stays always on the `strongly localized'
 side of the phase boundary. Note again the non-monotonic behavior. There is a pronounced maximum in the localization length close to $\lambda \approx 0.3$, which indicates that the system is least localized at this intermediate value of $\lambda$. Due to the nature of the approximation, $t(\lambda)$ and $\xi(\lambda)$ are underestimated, while the ratio $\xi(\lambda)/\xi(0)$ is overestimated. The value $\lambda_c\approx0.3$ extracted from the procedure in the Appendix, which includes additional approximations, is reasonably similar to $\lambda_c=0.5$ obtained by the simple argument detailed in the main text.}
\label{fig:analytic}
\end{center}
\end{figure}

\subsection{Numerics}

One way to numerically estimate the localization properties of the particle is by using numerical exact diagonalization and looking at the probability distribution of the position of the particle in the eigenstate $\ket{\Psi}$ of the coupled system and bath
\begin{equation}
p_i=\sum_{\alpha_i=1}^N|\bra{i,\alpha_i}\ket{\Psi}|^2.
\end{equation}
One can then define inverse participation ratios of $p$ as
\begin{equation}
I_q=\left(\sum_i p_i^{q}\right)^{-1}.
\end{equation}
For example the localization length can be estimated from the first non trivial $I_q$, i.e.\
\begin{equation}
I_2\sim\xi^d.
\end{equation}
Additional information is contained in the entropy of entanglement of system with bath, which may be extracted from the reduced density matrix $\rho=\Tr_{\text{bath}}\ket{\Psi}\bra{\Psi}$, where $\ket{\Psi}$ is an exact eigenstate of the coupled system and bath. The entanglement entropy is
\begin{equation}
S=-\Tr(\rho\ln\rho).
\end{equation}
For the present problem the entanglement entropy and the inverse participation ratios are correlated, the particle being more entangled with the bath the less localized it is.

\begin{figure}[htbp]
\begin{center}
\includegraphics[width=\columnwidth]{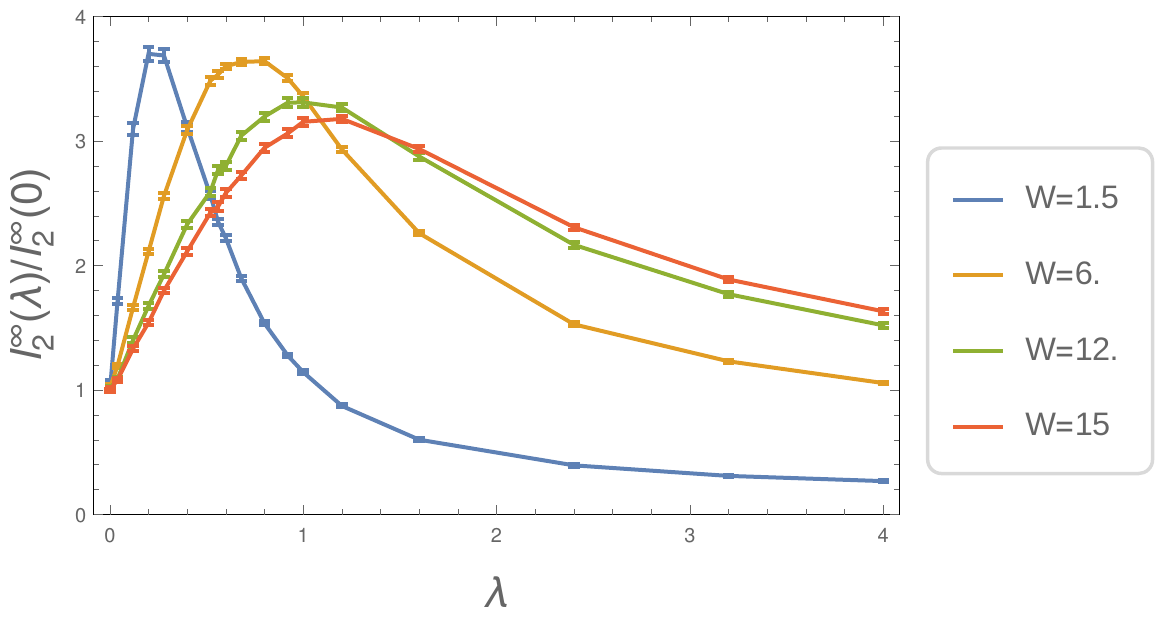}
\includegraphics[width=\columnwidth]{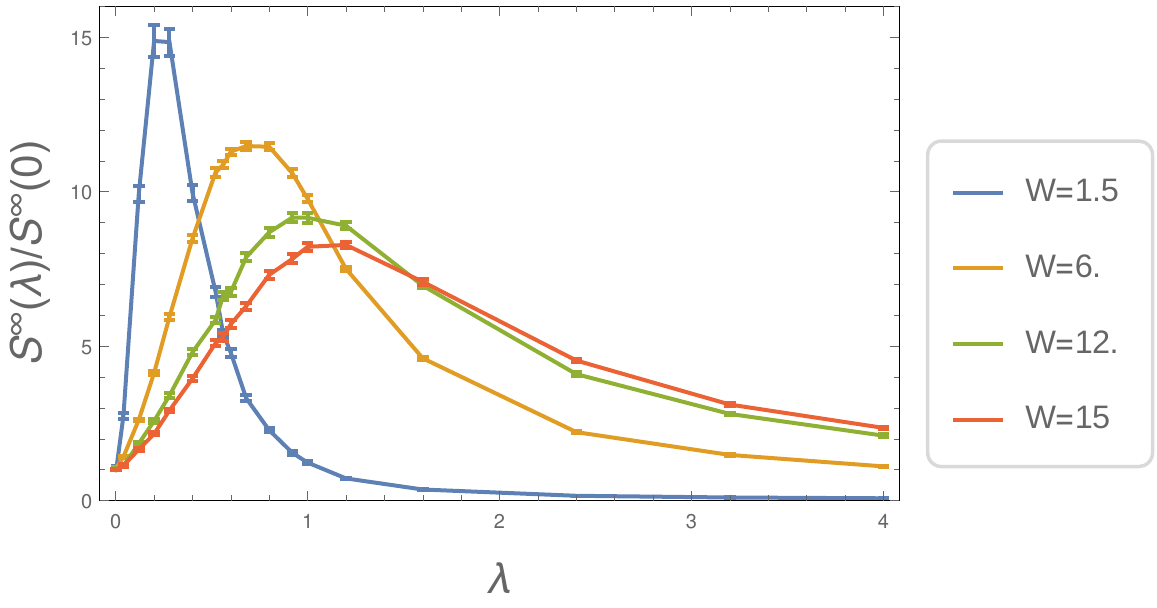}
\caption{Results of numerics on a one dimensional system coupled to a small bath. Exact diagonalization is performed for states in the center of the band for the parameters $t=1$, $\omega=4$, $N=300$, and varying $W$. The values of $W$ are chosen so that we span the phase diagram in Fig. \ref{fig:phdiag}. For $W < 12$, we slice through the `weak localization' region, whereas for $W > 12$ we stay always in the strong localization regime. The entanglement entropy and the participation ratios have a pronounced maximum close to the (same) hybridization threshold $\lambda_c$. The peak is sharper when we go through the weak localization regime. }
\label{fig:numeric}
\end{center}
\end{figure}

The numerical results for a one-dimensional system are presented in Fig. \ref{fig:numeric}, and show the evolution of the inverse participation ratio and entanglement entropy along horizontal slices taken through Fig. 2, some of which go through the `weak localization/delocalization' regime, while one does not. The numerics are for $t=1$, which is kept fixed, while the disorder $W$ varies.
Note that for $W\gtrsim 2 \sqrt{2N}/\omega\approx 12$ we are entirely in the strongly localized regime, for any value of $\lambda$. %For $W\gtrsim \left(\frac{2\sqrt{2N}}{\omega}\right)^\frac{1}{3}\approx 2.3$ we are entirely in the localized regime, for any value of $\lambda$. \addRN{Where do you get this estimate? Since the simulations are done for $t=1$ I would say we stay always on the strongly localized side of the phase boundary if $t_c(\lambda) > 1$ for all $\lambda$. Since $t_c$ is minimized at $\lambda = \lambda_c$, the condition then becomes $t_c(\lambda_c) = \sqrt{\omega W / 2 \sqrt{2N}} > 1$, or $W > 2 \sqrt{2N}/\omega$. This is different to the condition you write down. Also for the given parameters, it suggests that {\it all} of the curves pass through the WL regime, and we would need $W > 13$ to stay always on the localized side of the phase transition. Do you agree? And if yes, then could we do e.g. $W=15$? (if not, then lets note this).}
The first panel of Fig. \ref{fig:numeric} shows how the inverse participation ratio $I_2$ varies for an infinite size system coupled to a bath with coupling $\lambda$. We have extrapolated our finite size numerics to infinite $L$ by using a fit of the form
\begin{equation}
I_q(L)=I_q(\infty)+a/L,
\end{equation}
which turns out to be a very good fitting form for the data. We observe as expected a non-monotonic behavior, with weak $\lambda$ increasing the inverse participation ratio and strong $\lambda$ suppressing it.

The second panel on Fig. \ref{fig:numeric} shows the evolution of the entanglement entropy with coupling $\lambda$. At weak $\lambda$, the particle becomes more entangled with the bath as $\lambda$ is increased, but for larger $\lambda$ the entanglement entropy becomes a decreasing function of the coupling, and in the extreme $\lambda \rightarrow \infty$ limit one recovers an unentangled product state. The entanglement entropy is maximized at the same value of $\lambda = \lambda_c$ that maximizes the inverse participation ratio.

\begin{figure}[htbp]
\begin{center}
\includegraphics[width=\columnwidth]{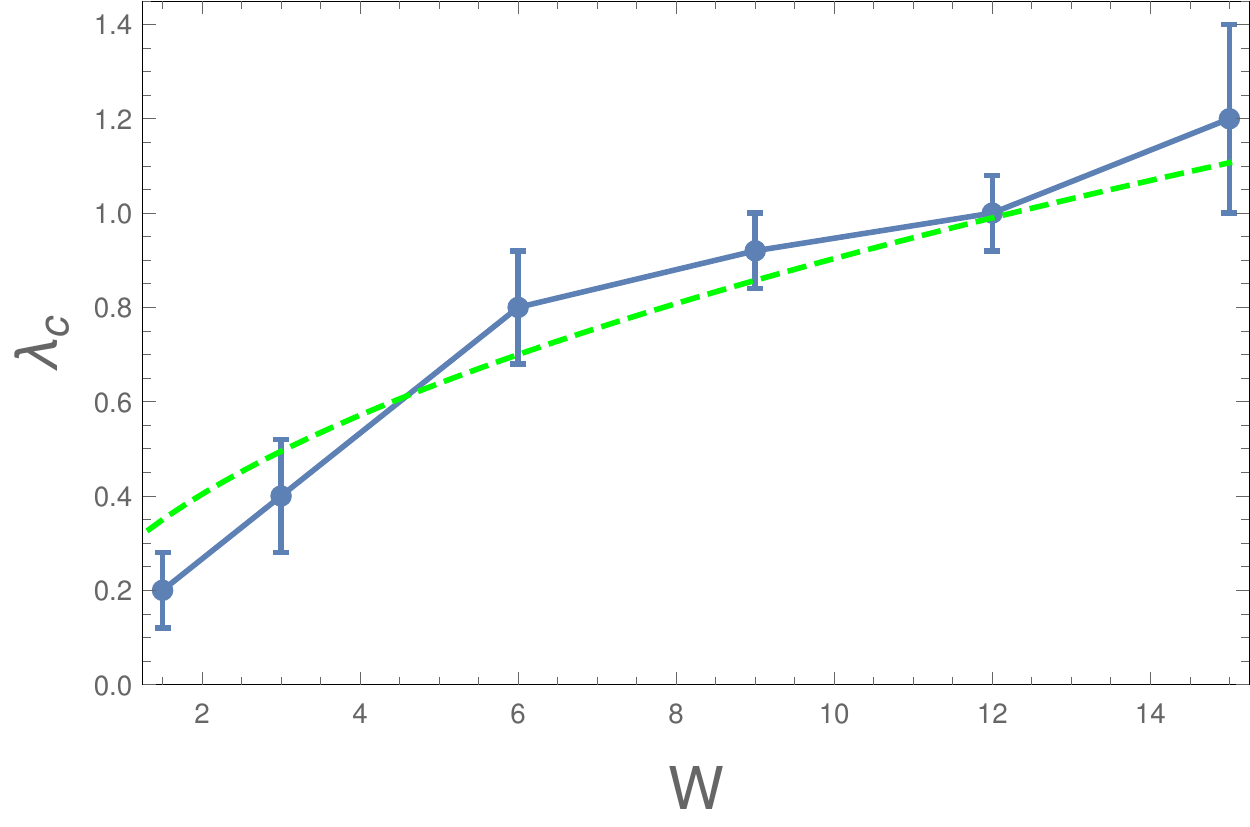}
\caption{Value of $\lambda_c$ for $t=1$ and the different disorders $W$. In blue we plot the numerical data, with their errors coming from the resolution in $\lambda$ at which the plots in Fig. \ref{fig:numeric} are computed. In dashed green we plot the analytic estimate of Eq. \eqref{lambdac}. We note that the exact diagonalization data agree well with the analytic prediction. }
\label{fig:lambdacrit}
\end{center}
\end{figure}

Fig. \ref{fig:lambdacrit} shows how the value of $\lambda_c$, which maximizes both the participation ratio and the entanglement entropy, varies with $W$. We compare the numerical result with our analytic estimate of Eq. \eqref{lambdac}, obtaining a good agreement.

\section{Outside the strong localization regime}
We now discuss what happens when the locator expansion fails to converge. What happens outside the strong localization regime is highly sensitive to dimensionality. In three dimensions this `strong hopping' regime will be delocalized, whereas in one or two dimensions it will be (weakly) localized, with a large localization length. This follows because even the strong hopping problem (schematically illustrated in Fig. 1) may be viewed as a (multi band) problem of a fermion moving in a random potential, in the orthogonal symmetry class, and this problem always displays localization in one and two dimensions \cite{gangoffour}. Indeed, one can even use the arguments of \cite{gangoffour} to estimate the localization length: the number of open conduction channels (loosely identified with the bare conductance $g_0$), is given by the ratio of matrix element to level spacing. This parameter takes value $g_0 \approx t/W$ for $\lambda < \omega/\sqrt{N}$, $g_0 \approx t \lambda \sqrt{N}/W \omega$ for $\omega / \sqrt{N} < \lambda < \lambda_c$, and $t/\lambda$ for $\lambda > \lambda_c$ according to the analysis we have just developed. Meanwhile, the typical conductance obeys the scaling relation
\begin{equation}
\frac{d \ln g}{d \ln L} = d-2 - \frac{c}{g}
\end{equation}
where $c$ is an unknown positive constant. In $d = 1,2$, $g$ always flows to zero, indicating localization. The localization length is simply the length scale on which $g$ becomes of order one. In one dimension this happens on length scales that are only power law large in $g_0$, whereas in two dimensions the localization length will be exponentially large in $g_0$. In three dimensions, meanwhile, the scaling function will flow to large $g$ for sufficiently large $g_0$, indicating the existence of a delocalized phase in the strong hopping regime. Thus, in three dimensions the `phase boundary' shown in Fig. 2 and Fig. 3, panel 1 is a true phase boundary separating localized and delocalized phases, whereas in one and two dimensions it merely marks a crossover from a strongly localized regime to a regime of weak localization.

%We note that in the weakly localized/delocalized regime the particle and the bath can be highly entangled. This follows because the exact eigenstates are superpositions of multiple product states. A particle on site $i$ can couple to multiple bath states, and moreover the particle can sit on any of $\min(\xi, L)^d$ sites. As a result, the exact eigenstates are superpositions of $\sim \min(L, \xi)^d$ product states. The entropy of entanglement of the system and bath is the log of the number of states in the superposition i.e.
%$S \sim d \ln \min(\xi, L)$. We note that for a thermodynamically large three dimensional system the divergence of the critical length defines a critical exponent $\xi\sim|W-W_c|^{-\nu}$ ($\nu=1$ in the approximation used in the Appendix) and thus at the localization-delocalization transition the entanglement entropy will diverge logarithmically $S\propto-\ln|W-W_c|$.

\section{Conclusions} Thus, we have examined the behavior of a single particle localized system coupled to a finite sized bath that is protected against localization. We find that the stability of localization is  a non-monotonic function of the coupling to the bath (Fig. 2).
%At the weakest couplings, the bath provides an additional source of static disorder, strengthening localization. 
At weak coupling, the bath weakens localization by placing hops in the system on shell.  At large coupling, the bath once again localizes the particle, by suppressing hopping through a mechanism akin to the quantum Zeno effect.  We have confirmed these results through detailed calculations in the forward approximation (Appendix) and by numerical exact diagonalization.

Our analysis has focused on a system containing a single particle. A detailed specification and solution of the many particle problem coupled to a small bath would be an interesting topic for future work.
%We now comment on the generalization to the many particle problem. In the region $\lambda < \delta$, when the coupling to the bath is weaker than the level spacing in the bath, the physics should not be materially altered (similar conclusions were reached in Ref. \onlinecite{ngh, johri}). Additionally, the regime of `quantum-Zeno' localization should still survive. However, elsewhere in the phase diagram the existence of an infinite range interaction (mediated through the bath) would appear to preclude localization. Thus, we expect that as we tune $\lambda$ in the many particle problem, we will go through two phase transitions, first from an Anderson localized phase  to a delocalized phase (at $\lambda \sim \delta$), and then from a delocalized phase to a Zeno localized phase. A more careful consideration of the many particle problem would be a worthwhile topic for future work.

A. S.\ is in part supported by the NSF grant PHY-1005429. A. S.\ would like to thank B. Altshuler for discussions.

\section{Appendix}

In this Appendix we review the main approximation for the analytic calculations discussed in the text. We will follow an analysis similar to that of \cite{rms-IOM}. In particular, we will see how to \emph{lift} the first order perturbation theory discussed in the text to a more precise result about the existence of localized states.

We will study the localized phase of the particle by doing perturbation theory in the hopping $t$.  The role of the bath coupling $\lambda$ is to open more and more channels for the particle to change its quantum number $i\to i+1$. This will define an effective hopping parameter $\tau$ and an effective connectivity $\kappa$ for the particle. An important quantity will be effective disorder $\mathcal{W}$, which is the energy spread of the energies $\epsilon_i+E_{\alpha_i}$ of neighboring states.
For simplicity of notation, we restrict to the case of a single particle in one dimension, and remark what has to be adapted in case of higher dimension.

For an eigenstate $\ket{\Psi}$ localized at $i=0$ for $t=0$, to lowest order in $t$ it holds
\begin{eqnarray}
\ket{\Psi}&=&\ket{0}\otimes\ket{\alpha_{0}}+\sum_{\alpha_1}A_{1,\alpha_1}\ket{1}\otimes\ket{\alpha_{1}}\nonumber\\
&+&\sum_{\alpha_2}A_{2,\alpha_2}\ket{2}\otimes\ket{\alpha_{2}}+...\nonumber\\
&+&\sum_{\alpha_n}A_{n,\alpha_n}\ket{n}\otimes\ket{\alpha_{n}}+...
\end{eqnarray}

where the amplitudes $A_{n,\alpha_n}$ are of order $t^n$, and can be written as

\begin{equation}
\label{eq:Ann}
A_{n,\alpha_n}=\hspace{-0.5cm}\sum_{p\in\text{paths}(\alpha_0,\alpha_n)}A_p,
\end{equation}
where
\begin{equation}\label{eq:locator}
A_p=\prod_{i=1}^n\frac{t \langle \alpha_{i-1}| \alpha_i \rangle}{\epsilon_i+E_{\alpha_i}-\epsilon_0-E_{\alpha_0}}
\end{equation}
is the amplitude of one particular path from $\alpha_0\to\alpha_n$, described by a particular sequence of bath states $p=(\alpha_1,\alpha_2,...,\alpha_n)$.

The amplitude for the particle to be at site $n$ equals $ \sum_{\alpha_n} A_{n,\alpha_n}$. In the following, we determine the probability to have a significant amplitude at a big distance $n$ from the localization center $i=0$ of the unperturbed eigenstate. More precisely, we identify the region of parameters where the probability to have a big amplitude converges to $0$ for $n \to \infty$, i.e. where
\begin{equation}\label{eq:loccond}
P\left( \left|\sum_{\alpha_n}A_{n,\alpha_n} \right|>z^n \right) \stackrel{n \to \infty}{\longrightarrow} 0
\end{equation}
for some $z<1$. In this regime of parameters, the locator expansion \eqref{eq:locator} converges and the system is strongly localized. The localization length $\xi$ is related to the minimum value $z_{\text{min}}$ for which \eqref{eq:loccond} holds by
\begin{equation}
\xi=\frac{1}{\ln z_{\text{min}}}\simeq\frac{1}{1-z_{\text{min}}},
\end{equation}
where the second (approximate) equality is only accurate close to the critical point, giving a mean-field exponent $\nu=1$. Setting $z=1$ in \eqref{eq:loccond}, we obtain an estimate of the critical hopping $t$, which for a given value of the parameters gives the breakdown of the locator expansion. Note that by considering only the lowest order in $t$ of $A_{n,\alpha_n}$ we are putting a lower bound on the critical hopping. This is because resonances have a much larger effect on the lowest order result than on the full result (here a `resonance' is a situation where $A_p$ is order one). Re-summation of higher order terms in $t$ decreases the final amplitude, so that the lowest order in $t$ calculation  overestimates the delocalizing effect of resonances. This was already discussed in the original paper by Anderson \cite{Anderson} (see also Ref. \onlinecite{de2014anderson}).

In the following, we compute the distribution of the amplitude $A_p$ of a particular  path from the state $\ket{i=0}\ket{\alpha_0}$ to the state $\ket{i=n}\ket{\alpha_n}$. As we shall see, for large $n$ the distribution is long tailed. As a consequence, both the sums \eqref{eq:Ann} and $ \sum_{\alpha_n} A_{n,\alpha_n}$ are very well approximated by their maximum term, and therefore
\begin{equation}
\left| \sum_{\alpha_n} A_{n,\alpha_n} \right| \simeq |\max_{\alpha_n}A_{n,\alpha_n}| \simeq\max_{\alpha_n}\max_{p\in\text{paths}(\alpha_0,...,\alpha_n)}|A_{p}|,
\end{equation}
which is effectively the maximum over all the paths of length $n$ emanating from $\alpha_0$, irrespective of the final state of the bath $\alpha_n$. We call this set of paths $\text{paths}^*(\alpha_0)$. Since each bath state $|\alpha_i \rangle$ can be chosen among $N$ possible states, the size of $\text{paths}^*(\alpha_0)$ is $N^n$.  Treating the different paths as independent we get:
\begin{equation}\label{eq:tu}
 P\left( \max_{p\in\text{paths}^*(\alpha_0)}|A_{p}|<z^n \right)=\left[ 1-   P\left( |A_{p}|>z^n \right) \right]^{N^n}.
\end{equation}
For a particle in a higher dimensional lattice, the sum \eqref{eq:Ann} has to be modified to account for the fact that two lattice sites at distance $n$ can be connected by multiple lattice paths of shortest length $n$, whose number is approximately equal to $d^{n}$.  For each of these paths the sequence of bath states can be chosen among $N^n$ possibilities. Thus \eqref{eq:tu} remains valid with the substitution $N \to N d$.
%\begin{equation}
% P\left( \max_{p\in\text{paths}^*(\alpha_0)}|A_{p}|<z^n \right)= e^{-N^n \log \left( P\left(|A_{p}|>z^n \right)\right)}.
%\end{equation}

We perform the computation of the distribution of \eqref{eq:locator} taking into account the dependence of the hopping amplitude on the energy difference between bath states which arises in the hybridized, intermediate region. In the intermediate regime we need to consider the hybridization of the bath eigenstates non-perturbatively. We can do this as follows: consider an eigenstate $\ket{\alpha_i}$ of $H_0=(\omega M^{(0)}+\lambda M^{(i)})$
\begin{equation}
H_0\ket{\alpha_i}=E_{\alpha_i}\ket{\alpha_i},
\end{equation}
and turn on the ``perturbation" $\lambda (M^{(i+1)}-M^{(i)})\equiv \lambda V$. Considering the new $H=H_0+\lambda V=\omega M^{(0)}+\lambda M^{(i+1)}$ with eigenstates $\ket{\alpha_{i+1}}$ we have on one hand the spectral decomposition
\begin{equation}
\label{eq:propagator}
\bra{\alpha_i}\frac{1}{E-H}\ket{\alpha_i}=\int dE_{\alpha_{i+1}}\rho(E_{\alpha_{i+1}})\frac{1}{E-E_{\alpha_{i+1}}}|\bra{\alpha_i}\ket{\alpha_{i+1}}|^2,
\end{equation}
on the other hand
\begin{equation}
\bra{\alpha_i}\frac{1}{E-H}\ket{\alpha_i}=\frac{1}{E-E_{\alpha_i}-\Sigma_{\alpha_i}(E)},
\end{equation}
where $\Sigma$ is the self-energy function.
We now take $E\to E_{\alpha_{i+1}}+i0^+$, and take the $\Im$ part of \eqref{eq:propagator} which gives:
\begin{equation}
\pi\rho(E_{\alpha_{i+1}})|\bra{\alpha_i}\ket{\alpha_{i+1}}|^2=\frac{\Delta}{(E_{\alpha_{i+1}}-E_{\alpha_i})^2+\Delta^2},
\end{equation}
where
\begin{equation}
\Delta=\Im\Sigma_{\alpha_i}(E_{\alpha_{i+1}}).
\end{equation}
Working at second order in the perturbation we have
\begin{equation}
\Sigma_{\alpha_i}(E)=\lambda^2\int dE_\beta\rho(E_\beta)\frac{|V_{\alpha,\beta}|^2}{E-E_\beta},
\end{equation}
and so
\begin{equation}
\Delta=\Im\Sigma_{\alpha_i}(E_{\alpha_{i+1}})=2\lambda^2\pi\rho(E_{\alpha_{i+1}})
\end{equation}
where we have assumed that $\bra{\alpha_i}\ket{\alpha_{i+1}}$ is some smooth function of the energy (true on average) and that $|V_{\alpha,\beta}|^2=2$, again true on average (we are in the region of well hybridized bath states, so average and typical values are the same).  Now recalling that $\rho(E)\equiv1/\delta$ we obtain
\begin{equation}
P(E',E)=|\bra{\alpha_i}\ket{\alpha_{i+1}}|^2=\frac{\Delta\delta/\pi}{(E-E')^2+\Delta^2},
\end{equation}
a result that we made use of in the main text.

To get the distribution of $|A_p|$, it is convenient to extract all energy scales and write the amplitude in terms of a new variable $Z_n>0$
\begin{equation}\label{eq:rescaling}
|A_p|=\left(\frac{t}{W} \sqrt{\frac{\delta}{ \pi \Delta }}\right)^n e^{Z_n},
\end{equation}
and then write the Laplace transform of the probability distribution $P_n(Z)$ of the variable $Z_n$:
\begin{equation}
g(s)=\int_0^\infty dZ P_{n}(Z) e^{-sZ}.
\end{equation}

Assuming $\epsilon_0 +E_{\alpha_0}=0$ (this gives the transition at the center of the band, or equivalently at $T=\infty$), we find

% We also assume that  $\Omega \sim \omega \sqrt{N} \gg W,\Delta$.

\begin{equation}
g(s)=\frac{1}{(s+1)^n}\left(\frac{8W}{\pi\Omega}\right)^{n}\tilde g(s),
\end{equation}
where
\begin{equation}\label{tilde}
\begin{split}
 \tilde{g}(s)=&\prod_{i=1}^n \frac{\Delta}{4 W} \int_{-\frac{\Omega}{2 \Delta}}^{\frac{\Omega}{2 \Delta}}\\
 &dE_i \sqrt{1 -\left(\frac{2 \Delta}{\Omega} E_i\right)^2} \Xi_{\frac{W}{\Delta}}( E_i,s) e^{\frac{s}{2} \log \left(1 +(E_i-E_{i-1})^2\right)}
 \end{split}
\end{equation}
with
\begin{equation}
\Xi_w(x,s)=\left|1+\frac{x}{w} \right|^{s+1} \mathrm{sgn} \left(w+x\right)+\left|1-\frac{x}{w}\right|^{s+1} \mathrm{sgn} \left(w-x\right)
\end{equation}
and $E_i\equiv E_{\alpha_i}/\Delta$ is dimensionless.

We need to invert the Laplace transform to get
\begin{equation}
P_{n}(Z)=\frac{1}{2 \pi i}\left(\frac{8 W}{\pi \Omega}\right)^n \int_{\mathcal{B}} ds \frac{e^{s Z} \tilde{g}(s)}{(s+1)^n}.
\end{equation}
 The Bromwich path is to the right of the $n$-pole $s=-1$. For the purpose of computing the large deviations giving rise to a resonance, we consider $Z=O(n)$ for large $n$, so $Z=n\zeta$. Then
\begin{equation}
P_{n}(Z)=\frac{1}{2 \pi i}\int_{\mathcal{B}} ds e^{n f(s)},
\end{equation}
with
 \begin{equation}\label{exp}
f(s)=s \zeta -\log (s+1) + \log \frac{8 W}{\pi \Omega}+\frac{1}{n}\log \tilde{g}(s)
 \end{equation}
can be computed with the saddle point expansion.

In the intermediate regime $\lambda< \omega$ it holds that $\Delta < \Omega$, thus we approximate $\sqrt{1 -\left(\frac{2 \Delta}{\Omega} E_i\right)^2}\to 1$; moreover, since the saddle point is dominated by the region with $s\to -1$, we can approximate $\Xi(x,s)\simeq \Xi(x,-1)=2\Theta(W/\Delta-|x|)$, so that
\begin{equation}\label{gtilde}
\begin{split}
 \tilde{g}(s)&\approx  \prod_{i=1}^n \frac{\Delta}{2 W} \int_{-\frac{W}{ \Delta}}^{\frac{ W}{\Delta}} dE_i  e^{\frac{s}{2} \log \tonde{1 +(E_i-E_{i-1})^2}}\\
 &=\tonde{ \prod_{i=1}^n \frac{\Delta}{2 W} \int_{-\frac{W}{ \Delta}}^{\frac{W}{\Delta}}dE_i }\\
 &\tonde{{1 +(E_n- E_{n-1})^2}}^{\frac{s}{2}}\tonde{{1 +(E_{n-1}- E_{n-2})^2}}^{\frac{s}{2}} \cdots\\
 &\quad \cdots \tonde{{1 +(E_2- E_{1})^2}}^{\frac{s}{2}}\tonde{{1 +E_1^2}}^{\frac{s}{2}}.
\end{split}
\end{equation}
This function is regular at $s=-1$ and it can be seen as the $n-$th application of an integral Kernel:
\begin{equation}
K(x',x)=\tonde{{1+q^2 (x'-x)^2}}^{\frac{s}{2}},
\end{equation}
with measure $d\mu(x)=dx/2$, to the function $\phi=(1+q^2x^2)^{s/2}$. Here $q=W/\Delta=   (\lambda_c /  \lambda )^2$, with $\lambda_c= \sqrt{W \omega/ 2 \sqrt{2 N}}$ introduced in \eqref{lambdac}.

%parameter $q=W/\Delta= (W \omega/  \sqrt{2N} \lambda^2)$, and that we work under the assumption $\omega/ \sqrt{N}<W< \omega \sqrt{N}$, the Anderson localized ($\lambda < \omega / \sqrt{N}$) and quantum Zeno regime  ($\lambda > \omega$) correspond to the large and small $q$ regime, respectively.
Denoting with $\alpha(s,q)$  the largest eigenvalue of $K$, we have, following the usual arguments for transfer matrix calculations
\begin{equation}
\tilde{g}(s)=c \; \alpha(s,q)^n
\end{equation}
to leading exponential order in $n$ ($c$ does not scale with $n$). As $K$ is a positive Kernel, by the Perron-Frobenius theorem, the largest eigenvalue is positive and corresponds to a positive eigenfunction $\mathcal{F}$ without any node on the interval $[-1,1]$ solving the equation
\begin{equation}
\label{eq:intker}
\int_{-1}^{1}\frac{dx}{2}(1+q^2(x'-x)^2)^{s/2}\mathcal{F}(x)=\alpha(s,q) \mathcal{F}(x').
\end{equation}

In the limit of small $q$ one has
\begin{equation}
K(x',x)\simeq1+\frac{sq^2}{2}(x'-x)^2+O(q^4).
\end{equation}
Then the eigenvalue problem can be solved exactly with an ansatz of the form $\mathcal{F}(x)=a+b x^2$ which for small $q$ gives
\begin{equation}
\label{eq:smallq}
\alpha=1+\frac{q^2 s}{3}+\frac{q^4 s^2}{45}+O(q^6),
\end{equation}
so to lowest order
\begin{equation}\label{eq:partsmallq}
\frac{1}{n}\log\tilde g (s)=q^2 s/3+O(q^4).
\end{equation}
 Inserting \eqref{eq:partsmallq} into \eqref{exp}, we get that $d f(s)/ds$ equals zero at the point
\begin{equation}
 s^*_{q \ll 1}=-1 +\frac{1}{\zeta + q^2/3},
\end{equation}
where
\begin{equation}
 f(s^*_{q \ll 1})= -\zeta +1 + \log \left(\frac{8 W}{\pi \Omega} (\zeta + q^2/3) \right)-\frac{q^2}{3}.
\end{equation}

Taking \eqref{eq:rescaling} into account we obtain:
\begin{equation}\label{eq:probsmallq}
P(|A_p|>z^n) \approx C_n \left[\frac{\sqrt{2} e }{\pi z}\frac{t}{\lambda N}  e^{-\frac{q^2}{3}} \log \left( \frac{2 z W}{t} \frac{\sqrt{N} \lambda}{\omega} e^{\frac{q^2}{3} } \right) \right]^n ,
\end{equation}
where $C_n$ scales sub-exponentially in $n$. Since this probability is exponentially small in $n$, from \eqref{eq:tu} one gets
\begin{equation}
  P\left( \max_{p\in\text{paths}^*(\alpha_0)}|A_{p}|>z^n \right) \approx1- e^{-N^n  P\left(|A_{p}|>z^n \right)},
\end{equation}
which approaches $0$ for increasing $n$ whenever $N P\left(|A_{p}|>z^n \right)^{1/n}<1$. Using \eqref{eq:probsmallq}, the condition for the critical hopping reads (in dimension $d$):
\begin{equation}\label{criterio1}
 \frac{\sqrt{2} e }{\pi}\frac{t d}{\lambda}  e^{-\frac{1}{24  } \left( \frac{W \omega}{\lambda^2 \sqrt{N}} \right)^2} \log \left(  \frac{2 W}{\omega} \frac{\lambda \sqrt{N}}{t} e^{\frac{1}{24  } \left( \frac{W \omega}{\lambda^2 \sqrt{N}} \right)^2} \right)=1,
\end{equation}
where we have taken the limit $z \to 1$. This is in agreement with the estimate \eqref{convergence} discussed in the main text.  Note that $W \omega/\lambda^2 \sqrt{N} = 2 \sqrt{2} (\lambda_c /\lambda)^2$ is small in the small $q$ regime that we are considering, and thus the exponential factors in \eqref{criterio1} can be neglected. Then the criterion for localization reduces to:
\begin{equation}
 \frac{\sqrt{2} e }{\pi}\frac{\tau \kappa}{\mathcal{W}}  \log \left(  \frac{2 W}{\omega \sqrt{N}} \frac{\mathcal{W}}{\tau}  \right)<1,
\end{equation}
which equals the critical condition for localization on a Bethe lattice with the effective parameters $\mathcal{W}= \lambda \sqrt{N}, \kappa=N d$ and $\tau=t/ \sqrt{N}$, up to an  additional factor  $W/ \omega \sqrt{N} \sim W/ \delta >1$ in  the logarithmic correction. Thus, the extrapolation to the Zeno regime is consistent with Eq. \eqref{eq:BL} in the main text.

For generic $q$ the integral equation is not easy to solve but the point $s=-1$ to which the saddle point is going to be very close, is regular. An approximation to the largest eigenvalue which works remarkably well for all values of $q$ and $s<0$ is obtained by taking the simple trial function $\mathcal{F}=1$.

\begin{figure}[htbp]
\begin{center}
\includegraphics[width=\columnwidth]{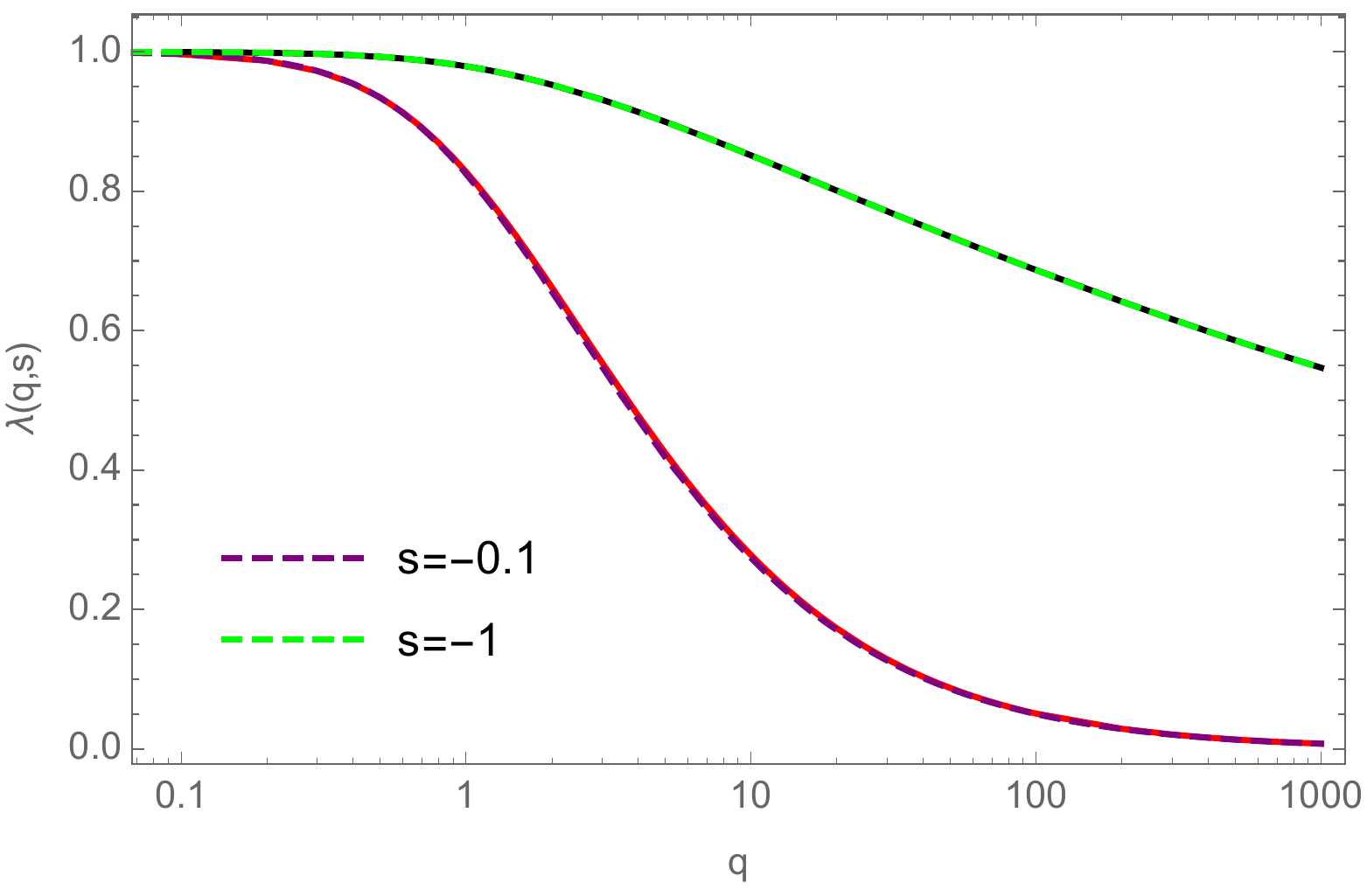}
\includegraphics[width=\columnwidth]{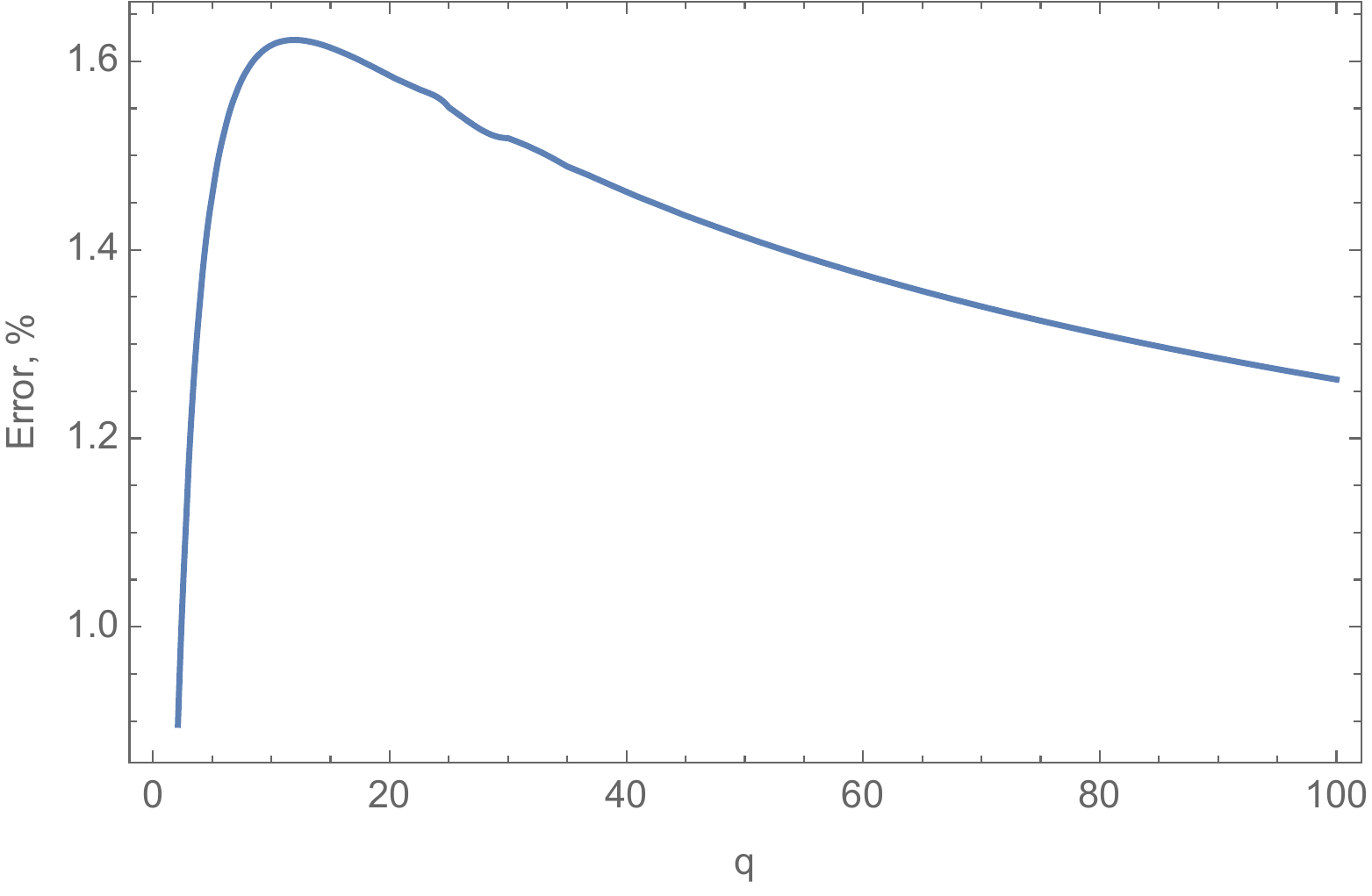}
\caption{Comparisons of the largest eigenvalue of the kernel in \eqref{eq:intker} and its approximate form \eqref{eq:applambda}. Above: Continuous line is the exact numerical results, dashed lines are analytical approximations. Below: Relative error, in percent for $s=-1$.}
\label{fig:LargestEV}
\end{center}
\end{figure}

The double integral $\alpha=\bra{\mathcal{F}}K\ket{\mathcal{F}}$ can then be transformed into a single integral (since it depends only on $x'-x$)
\begin{equation}
\alpha(q,s)=2\int_0^1(1-x)(1+q^2 x^2)^{s/2}dx
\end{equation}
and from this one gets:
\begin{equation}
\label{eq:applambda}
\begin{split}
\alpha(q,s)&=\frac{1}{2 q^2 (s+2)}\left(4 q^2 (s+2) F\left(\frac{1}{2},-\frac{s}{2};\frac{3}{2};-4 q^2\right)\right. \\
&-\left.\left(4 q^2+1\right)^{\frac{s}{2}+1}+1\right),
\end{split}
\end{equation}
where $F(a,b;c; x)$ is the hypergeometric function.

The comparison with the numerics is in Fig. \ref{fig:LargestEV}, the error never exceeds $1.6\%$ and is exact both in the large-$q$ and in the small-$q$ limit. For small $q$ one recovers \eqref{eq:smallq}.

 The approximate expression \eqref{eq:applambda} allows to derive an analytic estimate of the critical hopping in the regime of large $q$, where it holds
\begin{equation}
  \begin{split}
  \alpha(s,q)  &=   q^s  \tilde{\alpha}(s,q) + \mathcal{O} \left(\frac{1}{q^{2}}\right)
  \end{split}
  \end{equation}
  with
  \begin{equation}
   \tilde{\alpha}(s,q)=\left[ \frac{2^{s+1}}{(s+1)(s+2)}+\frac{1}{q^{1+s}} \frac{\sqrt{\pi} \Gamma (-(s+1)/2)}{2 \Gamma (-s/2) } \right].
  \end{equation}

 Neglecting higher order terms in $1/q$, the saddle point is attained at the point $s^*_{q \gg 1}$ satisfying:
   \begin{equation}
   s^*_{q \gg 1}=-1 + \frac{1}{\zeta+\log{q}+ \frac{d}{ds}\log \tilde{\alpha}(s^*_{q \gg 1},q)}.
  \end{equation}

The function $\log \tilde{\alpha}(s,q)$ has the expansion:
\begin{equation}
 \log \tilde{\alpha}(s,q)=\alpha_0(q) + \alpha_1(q) (s + 1)+ \mathcal{O}((s+1)^2),
\end{equation}
with
\begin{equation}
 \begin{split}
 \alpha_0(q)&=\log \left[\log (4q)-1\right],\\
 \alpha_1(q)&= -\frac{\log q}{2}-\frac{1}{2} + \mathcal{O} \left( \frac{1}{\log q} \right).
  \end{split}
\end{equation}

For $q$ large, the saddle point $ s^*_{q \gg 1}$ approaches the point $s=-1$; in this regime one can therefore set
 \begin{equation}\label{eq:saddleqlarge}
 \begin{split}
   s^*_{q \gg 1}&\approx -1 + \frac{1}{\zeta+\log{q}+ \frac{d}{ds}\log \tilde{\alpha}(-1,q)}\\
    &= -1 + \frac{1}{\zeta+\log{q}+ \alpha_1(q)}.\\
    \quad
          \end{split}
  \end{equation}

Substitution into \eqref{exp} gives:
\begin{equation}
 \begin{split}
  f(s^*_{q \gg 1})=&-\zeta+1+\log \left[ \zeta+\log q + \alpha_1(q)\right]+\\
  &+\log \left(\frac{8 W}{\pi \Omega } \frac{\log (4q)}{q}\right)+\mathcal{O} \left(\frac{1}{\log q}\right),
 \end{split}
\end{equation}
from which one gets
\begin{equation}
\begin{split}
 P(|A_p|>z^n) \approx& D_n \left[ \frac{4 e}{\pi z} \frac{t \lambda}{W \omega \sqrt{N}} \log \left( \frac{2 W \omega}{\lambda^2 \sqrt{2 N}} \right)\right]^n \cdot\\
 &\cdot\log^n \left( \frac{ W }{t} \sqrt{\frac{W \sqrt{2 N}}{\omega}} \right),
 \end{split}
\end{equation}
with $D_n$ scaling sub-exponentially with $n$. In this limit, the locator expansion converges (in $d$ dimensions) for:
\begin{equation}
 \frac{4 e}{\pi } \frac{t d}{W} \frac{ \lambda  \sqrt{N}}{ \omega} \log \left( \frac{2 W \omega}{\lambda^2 \sqrt{2 N}} \right)\log \left( \frac{ W }{t} \sqrt{\frac{W \sqrt{2 N}}{\omega}} \right)<1,
\end{equation}
which is in agreement with the estimate \eqref{convergence} in the main text.

For arbitrary $q$ the saddle point equation in $s$ has to be solved numerically.
The estimate of the critical value of the hopping is obtained solving numerically the equation
\begin{equation}
 N P_{n}(n \log( W(\Delta \pi)^{1/2}/t\delta^{1/2})) ^{1/n} =1
\end{equation}
which is equivalent to $N P\left(|A_{p}|>z^n \right)^{1/n}=1$ with $z \to 1$. The result for $d=1$ is plotted in Fig. \ref{fig:analytic} in the text.


\begin{thebibliography}{1}%
\makeatletter
\providecommand \@ifxundefined [1]{%
 \@ifx{#1\undefined}
}%
\providecommand \@ifnum [1]{%
 \ifnum #1\expandafter \@firstoftwo
 \else \expandafter \@secondoftwo
 \fi
}%
\providecommand \@ifx [1]{%
 \ifx #1\expandafter \@firstoftwo
 \else \expandafter \@secondoftwo
 \fi
}%
\providecommand \natexlab [1]{#1}%
\providecommand \enquote  [1]{``#1''}%
\providecommand \bibnamefont  [1]{#1}%
\providecommand \bibfnamefont [1]{#1}%
\providecommand \citenamefont [1]{#1}%
\providecommand \href@noop [0]{\@secondoftwo}%
\providecommand \href [0]{\begingroup \@sanitize@url \@href}%
\providecommand \@href[1]{\@@startlink{#1}\@@href}%
\providecommand \@@href[1]{\endgroup#1\@@endlink}%
\providecommand \@sanitize@url [0]{\catcode `\\12\catcode `\$12\catcode
  `\&12\catcode `\#12\catcode `\^12\catcode `\_12\catcode `\%12\relax}%
\providecommand \@@startlink[1]{}%
\providecommand \@@endlink[0]{}%
\providecommand \url  [0]{\begingroup\@sanitize@url \@url }%
\providecommand \@url [1]{\endgroup\@href {#1}{\urlprefix }}%
\providecommand \urlprefix  [0]{URL }%
\providecommand \Eprint [0]{\href }%
\providecommand \doibase [0]{http://dx.doi.org/}%
\providecommand \selectlanguage [0]{\@gobble}%
\providecommand \bibinfo  [0]{\@secondoftwo}%
\providecommand \bibfield  [0]{\@secondoftwo}%
\providecommand \translation [1]{[#1]}%
\providecommand \BibitemOpen [0]{}%
\providecommand \bibitemStop [0]{}%
\providecommand \bibitemNoStop [0]{.\EOS\space}%
\providecommand \EOS [0]{\spacefactor3000\relax}%
\providecommand \BibitemShut  [1]{\csname bibitem#1\endcsname}%
\let\auto@bib@innerbib\@empty
%</preamble>
\bibitem [{Note1()}]{Note1}%
  \BibitemOpen
  \bibinfo {note} {In the following we will consider states in the middle of
  the spectrum. Outside of the center of the band other phenomena arise. In
  particular due to the reduction of the density of states the localized phase
  is expected to be larger than the limits computed here.}\BibitemShut {Stop}%
\end{thebibliography}%


\begin{thebibliography}{99}
\bibitem[Anderson (1958)]{Anderson}
P. W. Anderson, Phys. Rev. {\bf 109}, 1492 (1958).
\bibitem{Fleishman}
L. Fleishman and P. W. Anderson, Phys. Rev. B 21, 2366 (1980).
\bibitem[AKGL (1997)]{agkl}
B. L. Altshuler, Y. Gefen, A. Kamenev and L. S. Levitov, Phys. Rev. Lett. {\bf 78}, 2803 (1997).
\bibitem[Mirlin (2006)]{Mirlin}
I. V. Gornyi, A. D. Mirlin and D. G. Polyakov, Phys. Rev. Lett. {\bf 95}, 206603 (2005).
\bibitem[BAA (2006)]{BAA}
D. M. Basko, I. L. Aleiner and B. L. Altshuler, Annals of Physics {\bf 321}, 1126 (2006).
\bibitem[Oganesyan (2008)]{Oganesyan}
V. Oganesyan and D. A. Huse,
Phys. Rev. B {\bf 75}, 155111 (2007).
\bibitem[Znid (2008)]{Znid}
M. Znidaric, T. Prosen and P. Prelovsek, Phys. Rev. B {\bf 77}, 064426 (2008).
\bibitem[Pal (2010)]{pal}
A. Pal and D. A. Huse,
Phys. Rev. B {\bf 82}, 174411 (2010).
\bibitem{lbits} D.A. Huse and V. Oganesyan, arXiv: 1305.4915, D.A. Huse, R. Nandkishore and V. Oganesyan, {\it Phys. Rev. B} {\bf 90}, 174202 (2014)
\bibitem{serbyn} M. Serbyn, Z. Papic and D. A. Abanin, {\it Phys. Rev. Lett.} {\bf 111}, 127201 (2013).
\bibitem{imbrie}
J. Z. Imbrie, arXiv:1403.7837.
\bibitem{rms-IOM} V.Ros, M.Mueller, A.Scardicchio, Nucl. Phys. B 891, pp. 420-465 (2015).
\bibitem{chandran} A. Chandran, I. H. Kim, G. Vidal, D. A. Abanin, Phys. Rev. B 91, 085425 (2015).
\bibitem{kim}  I. H. Kim, A. Chandran, D. A. Abanin,  arXiv: 1412.3073.
\bibitem{adiabaticity} V. Khemani, R. Nandkishore and S.L. Sondhi, arXiv: 1411.2616
\bibitem[LPQO (2013)]{LPQO}
D. A. Huse, R. Nandkishore, V. Oganesyan, A. Pal and S. L. Sondhi, Phys. Rev. B {\bf 88}, 014206 (2013).
\bibitem[Pekker (2013)]{Pekker}
D. Pekker, G. Refael, E. Altman, E. Demler and V. Oganesyan, Phys. Rev. X {\bf 4}, 011052 (2014).
\bibitem[Vosk (2013)]{Vosk}
R. Vosk and E. Altman, Phys. Rev. Lett. {\bf 112}, 217204 (2014).
\bibitem{Kjall}
J. A. Kjall, J. H. Bardarson and F. Pollmann, Phys. Rev. Lett. 113, 107204 (2014).
\bibitem[Bauer (2013)]{Bauer}
B. Bauer and C. Nayak, J. Stat. Mech. P09005 (2013).
\bibitem{deluca2013} A.\ De Luca, A.\ Scardicchio, EPL (Europhysics Letters), {\bf 101}, 37003 (2013).
\bibitem[Bahri (2013)]{Bahri}
Y. Bahri, R. Vosk, E. Altman and A. Vishwanath, arXiv:1307.4192.
\bibitem{lspt} A. Chandran, V. Khemani, C. R. Laumann and S. L. Sondhi, Phys. Rev. B {\bf 89}, 144201 (2014).
\bibitem{qhmbl} R. Nandkishore and A. C. Potter, {\it Phys. Rev. B} {\bf 90}, 195115 (2014)
\bibitem{arcmp}
R. Nandkishore and D. A. Huse, arXiv:1404.0686, Ann. Rev. Cond. Matt. Phys. [in press].
\bibitem{laumann2014many} C.R.\ Laumann, A. Pal, and A. Scardicchio, Phys. Rev. Lett. 113, 200405 (2014).
\bibitem{deluca2014Bethe} A. De Luca, B.L.\ Altshuler, V.E.\ Kravtsov, and A.\ Scardicchio
Phys. Rev. Lett. 113, 046806 (2014).
\bibitem[ngh (2014)]{ngh}
R. Nandkishore, S. Gopalakrishnan and D.A. Huse, {\it Phys. Rev. B} {\bf 90} 064203 (2014)
\bibitem[johri (2014)]{johri}
S. Johri, R. Nandkishore and R.N. Bhatt, arXiv: 1405.5515 (2014)
\bibitem[gn (2014)]{gn}
S. Gopalakrishnan and R. Nandkishore, arXiv: 1405.1036
\bibitem[2dcontinuum (2014)]{2dcontinuum}
R. Nandkishore, {\it Phys. Rev. B} {\bf 90}, 184204 (2014)
\bibitem[Chalker (2003)]{Chalker}
V. Gurarie and J.T. Chalker, Phys. Rev. B 68, 134207 (2003)
\bibitem[Thouless (2984)]{Thouless}
D.J. Thouless, J. Phys. C: Solid State Phys. 17, L325 (1984)
\bibitem[Mehta (1984)]{Mehta}
M.L. Mehta, {\it Random Matrices, $3^{rd}$ ed.} , Elsevier Inc., Amsterdam (2004)
\bibitem[gangoffour (1979)]{gangoffour}
E. Abrahams, P.W. Anderson, D.C. Licciardello and T.V. Ramakrishnan, Phys. Rev. Lett. {\bf 42}, 673 (1979)
\bibitem[DeLuca (2014)]{de2014anderson}
A. De Luca, B.L. Altshuler, V.E. Kravtsov and A. Scardicchio, Physical Review Letters {\bf 113}, 046806 (2014).
\bibitem[Beskow (1967)]{Beskow}
A. Beskow and J. Nilsson, Ark. Fys. {\bf 34}, 561 (1967)
\bibitem[Khalfin (1968)]{Khalfin}
L.A. Khalfin, JETP Lett. {\bf 8}, 65, (1968)
\bibitem[Misra (1977)]{Misra}
B. Misra and E.C.G. Sudarshan, J. Math. Phys. (N.Y.) {\bf 18}, 756 (1977)
\bibitem{Facchi2001}
P. Facchi, S. Pascazio, A. Scardicchio, and L. S. Schulman
Phys. Rev. A 65, 012108 (2001).
\bibitem[Facchi (2002)]{Facchi}
P. Facchi and S. Pascazio, Phys. Rev. Lett. {\bf 89}, 080401 (2002)
\end{thebibliography}
\end{document}